\documentclass[reprint,
showpacs,
 amsmath,amssymb,
 aps,
]{revtex4-1}

\usepackage{graphicx}
\usepackage{dcolumn}
\usepackage{bm}
\usepackage{amsmath}
\usepackage{dnaseq}
\usepackage{caption}
\usepackage{subcaption}
\newcommand{\beginsupplement}{%
        \setcounter{table}{0}
        \renewcommand{\thetable}{S\arabic{table}}%
        \setcounter{figure}{0}
        \renewcommand{\thefigure}{S\arabic{figure}}%
        \setcounter{equation}{0}
        \renewcommand{\theequation}{S\arabic{equation}}%
}

\begin{document}

\preprint{APS/123-QED}

\title{Probing the elastic limit of DNA bending}

\author{Tung T. Le}
\author{Harold D. Kim}%
 \email{harold.kim@physics.gatech.edu}
\affiliation{%
 School of Physics, Georgia Institute of Technology\\ 837 State Street, Atlanta, GA 30332\\ 
}%

\date{\today}
\begin{abstract}
Many structures inside the cell such as nucleosomes and protein-mediated DNA loops contain sharply bent double-stranded (ds) DNA. Therefore, the energetics of strong dsDNA bending constitutes an essential part of cellular thermodynamics. Although the thermomechanical behavior of long dsDNA is well described by the worm-like chain (WLC) model, the length limit of such elastic behavior remains controversial. To investigate the energetics of strong dsDNA bending, we measured the opening rate of small dsDNA loops with contour lengths of 40-200 bp using Fluorescence Resonance Energy Transfer (FRET). From the measured relationship of loop stability to loop size, we observed a transition between two separate bending regimes at a critical loop size below 100 bp. Above this loop size, the loop lifetime decreased with decreasing loop size in a manner consistent with an elastic bending stress. Below the critical loop size, however, the loop lifetime became less sensitive to loop size, indicative of softening of the double helix. The critical loop size was measured to be ${\sim}$60 bp with sodium only and ${\sim}$100 bp with 5 mM magnesium, which suggests that magnesium facilitates the softening transition. We show that our results are in quantitative agreement with the kinkable worm-like chain model. Furthermore, the model parameters constrained by our data can reproduce previously measured J factors between 50 and 200 bp. Our work provides powerful means to study dsDNA bending in the strong bending regime. 

\end{abstract}


                            
\maketitle


\section{\label{intro}Introduction}

Double-stranded DNA (dsDNA) can bend, twist, stretch, and adopt various structures under thermal excitation\cite{olson2000modeling,strick2000twisting,nikolova2011transient,maehigashi2012b}. Despite this wide range of thermal fluctuations, mechanical properties of DNA at large length scales can be well-described by the worm-like chain (WLC) model \cite{shimada1984ring}. According to this model, directional change in the chain contour costs energy quadratically dependent on bending angle. Chain stiffness can be described by the persistence length below which thermally induced bending fluctuation becomes negligible. The persistence length of dsDNA has been estimated to be 45-50 nm by various methods\cite{elias1981transient,bednar1995determination,smith1996overstretching,bouchiat1999estimating,shore1981dna,taylor1990application}, and shown to be largely independent of monovalent salt concentration above 20 mM\cite{porschke1991persistence,baumann1997ionic,wenner2002salt}.

Strong bending of dsDNA, which refers to deflection of larger than ${\sim}2.4\,^{\circ}$ between adjacent base pairs (or equivalently, one turn per persistence length), occurs in transcriptional repression\cite{bond2010gene}, nucleosome formation\cite{andrews2011nucleosome}, and viral DNA packaging\cite{baker1999adding}. Since the WLC is valid only within the elastic limit of dsDNA, the actual bending energy of dsDNA in such processes may deviate from the WLC prediction. The free energy cost of dsDNA bending can be experimentally determined by measuring the efficiency with which a linear dsDNA can be ligated into a circle. By comparing the rates of circle and dimer formation in the ligation reaction, one can obtain an effective molar concentration of one end of the DNA around the other end, which is known as the J factor\cite{shore1981dna}.  Using this ligase-dependent cyclization assay, the Widom group showed that the J factors of dsDNAs shorter than 150 bp were several orders of magnitude higher than the WLC predictions\cite{cloutier2004spontaneous,cloutier2005dna}. 

Subsequently, several other groups used different experimental methods to draw similar conclusions that dsDNA bends more readily than predicted\cite{wiggins2006high,yuan2008dna,han2009concentration}. To explain the apparent failure of the WLC model, structural inhomogeneities such as bubbles or kinks have been proposed as mechanisms for enhanced flexibility in the strong bending regime\cite{yan2004localized,wiggins2005exact}. The meltable or kinkable WLC model could correctly predict the measured J factors, but the parameters used are not supported by experimental data\cite{sivak2012consequences}.

Du et al. later pointed out that the ligase concentration used in the first study by the Widom group was too high to correctly estimate the J factor\cite{du2005cyclization}. They measured the bimolecular rate constant of dimerization in a separate ligation reaction using low ligase concentration, and showed that the measured J factor is in agreement with the WLC model. The other experimental studies that reported high J factors used AFM on surface-confined DNA and tethered particle motion on protein-mediated DNA looping. These techniques can bias the equilibrium looping probability distribution due to surface interaction\cite{destainville2009microscopic}, nonspecific binding of proteins to DNA\cite{yuan2007t4,manzo2012effect} or the presence of a bead\cite{han2009concentration}. 

As an alternative method free of these concerns, single-molecule FRET has recently been used to measure the J factors of short dsDNAs\cite{vafabakhsh2012extreme}. In this method, DNA loop formation can be detected using FRET without the need to use external agents. Vafabakhsh and Ha found that J factors in the range between 65 and 110 bp determined from looping kinetics were a few orders of magnitude higher than the WLC model prediction. The results from this study suggest a significant departure of dsDNA from either the WLC model or 45-50-nm persistence length. However, other experimental factors could have led to an overestimation of the J factor: (1) using synthetic oligos may introduce mismatched base pairs\cite{vologodskii2013strong}, (2) high salt conditions (1 M sodium or 10 mM magnesium) can increase DNA curvature and flexibility\cite{brukner1994physiological,strauss1994dna,williams2000electrostatic,stellwagen2010effect} and/or (3) long sticky ends used in the experiment can increase the apparent looping probability\cite{peters2010dna,vologodskii2013bending}. (1) can be addressed by using PCR-based DNA assembly\cite{vologodskii2013strong,le2013measuring}, but (2) and (3) cannot be easily addressed because lowering salt concentration or shortening the sticky ends severely reduce the frequency of looping events observable by FRET for short DNA molecules.

\begin{figure}[htp]
\includegraphics[width=\columnwidth]{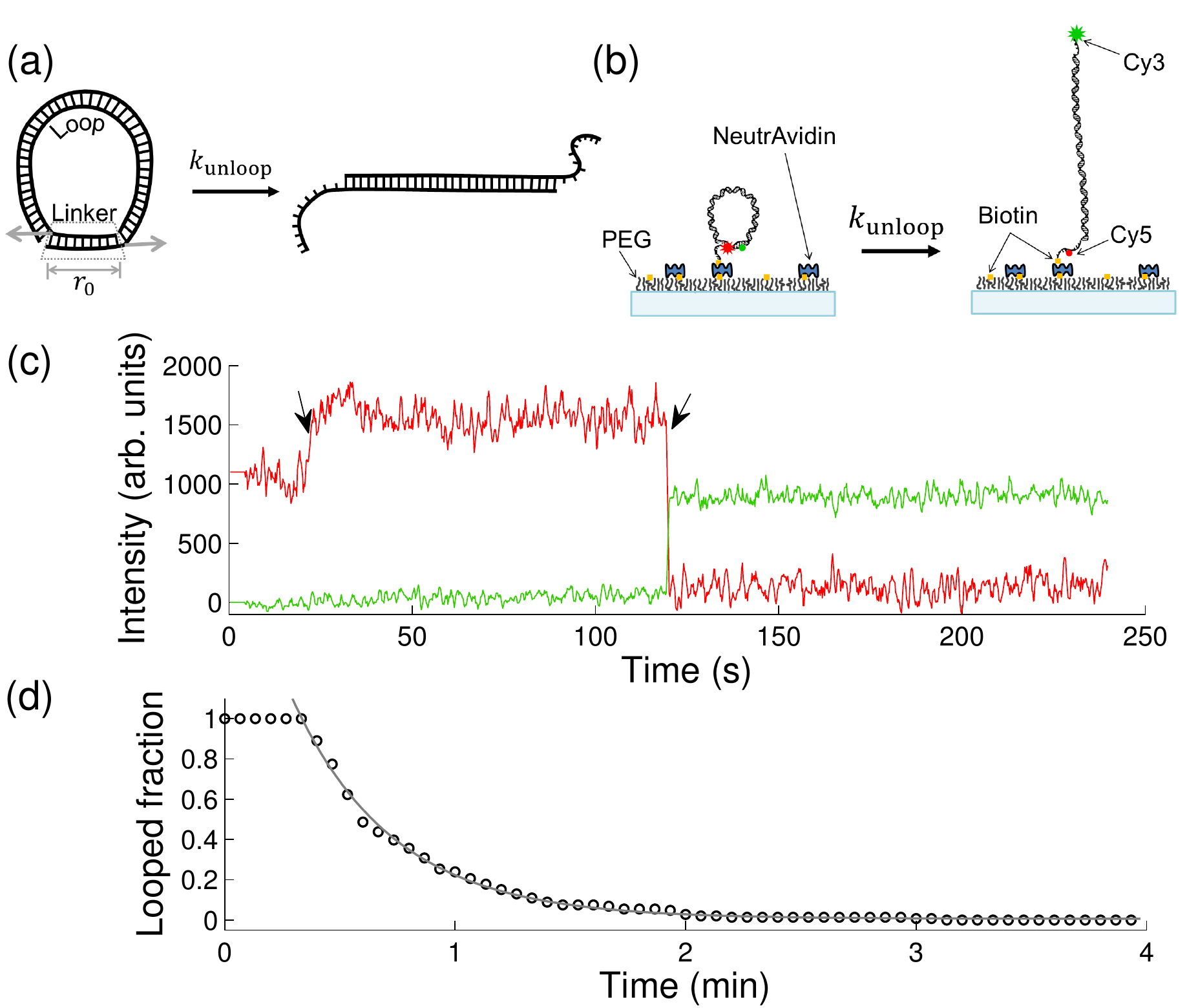}
\caption{\label{fig1}{\bf Loop breakage assay.} (a) The shear force exerted on the linker duplex by the loop.  The force exerted in the shear direction (gray arrows) accelerates dissociation of the linker duplex according to the Bell relationship (Eq. ˜\ref{bell}). (b) DNA design. A short DNA with sticky ends can be captured in the looped state when the two sticky ends are annealed. The looped state (left) and the unlooped state (right) correspond to high FRET and low FRET, respectively. For single-molecule experiments, DNA molecules are immobilized through a biotinylated end to a PEG-coated glass surface. In [Na\textsuperscript{+}] = 2 M, a significant fraction of molecules exist in the looped state. Decreasing [NaCl] from 2 M to 50 mM by flow induces breakage of DNA loops. (c) A representative time trace of Cy3 (green) and Cy5 (red) intensities from a single molecule. The change in salt concentration causes an increase in the Cy5 intensity due to an unknown reason (marked by a black arrow). Upon loop breakage, Cy5 intensity drops, and Cy3 intensity jumps (marked by a black arrow). (d) The time decay of the number of dsDNA loops upon salt concentration drop. The molecules begin to unloop shortly after perfusion of 50 mM [Na\textsuperscript{+}] buffer. The decay curve is fitted with a single exponential function to extract the lifetime of the DNA loop.}
\end{figure}
In this paper, we take a different FRET-based approach to test the WLC model at short length scales. The key idea is that stability of end-to-end annealed DNA loops is highly sensitive to loop size due to internal bending stress as depicted in FIG. ~\ref{fig1}(a). In our FRET assay, the looped state of a dsDNA is stabilized by formation of a transient linker duplex of ${\sim}$10 bp between its sticky ends. The lifetime of this linker duplex depends on the shear force exerted along its helical axis by the looped DNA. Since different DNA models make different predictions about how this shear force depends on the loop length, we can experimentally test these models by measuring linker lifetime vs. loop size. 

Our unlooping-based approach has unique capabilities that complement the ligation-based or FRET-based J factor measurements: (1) unlooping rates can be measured with good statistics in moderate salt conditions where looping of short dsDNA rarely occurs; (2) only the molecules that were able to loop are followed in the loop breakage assay, which automatically filters out dysfunctional molecules; and (3)  the unlooping rate is related to the shear force, which is easier to compute than the J factor.

Using this unlooping assay, we measured the lifetime of small DNA loops as a function of loop size in the strong bending regime. We found that the loop lifetime decreases with decreasing loop size, indicative of increasing bending stress. The bending stress, however, ceased to increase elastically below a critical loop size, reminiscent of a structural transition in dsDNA, such as kink formation. Based on this apparent transition, we estimate the free energy of kink formation to be larger than ${\sim}18 k_B T$. We also found that this energy cost was significantly lowered by magnesium to ${\sim}12 k_B T$. Based on our findings, we propose a kinkable worm-like chain (KWLC) model with salt-dependent kinkability to resolve the apparent discrepancy between previous J factor measurements. 

\section{\label{res}Results}
DNA molecules with sticky ends were constructed using a PCR-based protocol\cite{le2013measuring}. Cy3 and Cy5, the donor-acceptor pair for FRET are incorporated near the sticky ends of the DNA so that loop stabilization by the sticky ends results in high FRET efficiency. A biotin linker extends from one end for surface immobilization (FIG. ~\ref{fig1}(b)). The DNA sequences used in this study are random and do not contain A-tracts which can produce curved molecules. The DNA molecules immobilized to the surface are first stabilized in the looped state in a buffer with 2 M [Na\textsuperscript{+}]\cite{vafabakhsh2012extreme}. Once equilibrium is reached, an imaging buffer containing 50-200 mM [Na\textsuperscript{+}] is perfused into the sample chamber, and Cy3 and Cy5 fluorescence intensities are continuously monitored (FIG. ~\ref{fig1}(c)). The number of remaining high-FRET molecules is recorded as a function of time, and the decay curve is fitted with a single exponential function to extract the linker lifetime (FIG. ~\ref{fig1}(d)). 

\begin{figure}[htp]
\includegraphics[width=\columnwidth]{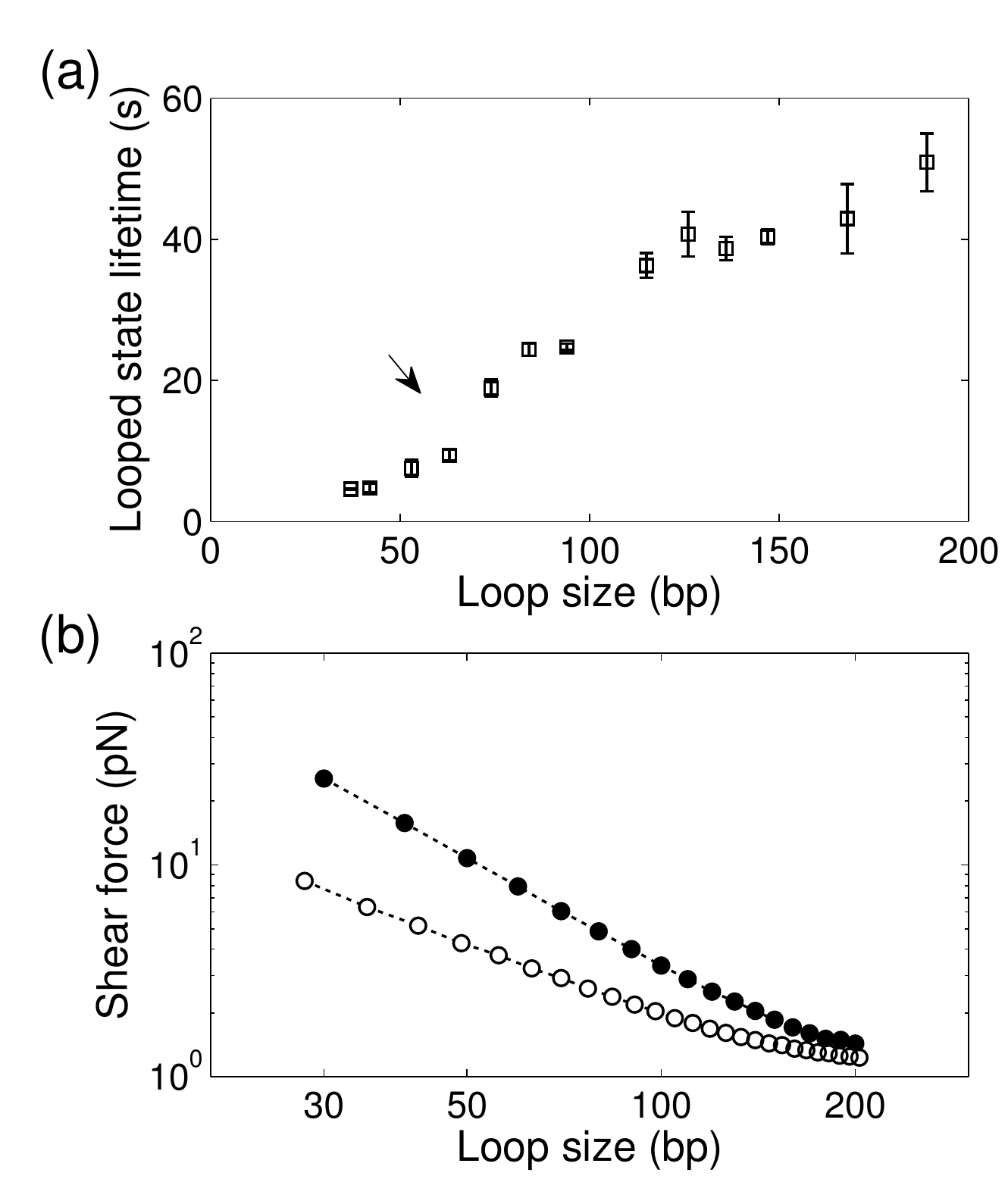}
\caption{\label{fig2}{\bf Shear-force dependent loop breakage.} (a) Looped state lifetime vs. loop size. The loop lifetime at various size was measured in 50 mM NaCl. The error bar is the standard error of the mean from at least 4 measurements. The black arrow indicates the inflection point. (b) Line-scatter log-log plot of calculated shear force vs. loop size. The shear force is calculated from the biased Monte Carlo (MC) simulation using the WLC model with a persistence length of 50 nm (filled circles) or the LSEC model (open circles). For each loop size, we performed three simulations, each with ${\sim}5\times10^6$ accepted conformations. The errorbar size is typically smaller than the size of the symbol. See Methods for more information on the simulation details.}
\end{figure}

We repeated this salt drop experiment for different lengths of DNA molecules ranging from 40 to 200 bp. In this length range, the bending energy dominates the free energy of looping. Since the total bending energy of the loop increases as the loop size decreases (Supplementary Information and FIG. ~\ref{sfig1}), we expect smaller loops to become less stable. In support of this notion, the linker lifetime decreased as the DNA length was reduced (FIG. ~\ref{fig2}(a)). Interestingly, the curve exhibits inflection near 70 bp from concave up to concave down. 

To gain more insights into this apparent inflection, we formulate the relationship between the lifetime and the loop size by using the shear force exerted on the linker duplex as an intermediate variable. The lifetime ($\tau$) of the linker duplex of length $r_0$ subjected to a shear force ($f$) can be modeled by the Bell relationship\cite{bell1978models,bustamante2004mechanical}
\begin{subequations}
 \label{bell}
 \begin{eqnarray}
 \tau (f) = \tau (0) \exp \left ( -\frac{f\Delta r_0}{k_B T} \right ), \label{eqa}\\
 \log \tau (f) = \log \tau (0) - \frac{f\Delta r_0}{k_B T}, \label{eqb}
 \end{eqnarray}
\end{subequations}
where $\Delta r_0$ is the elongation of the linker duplex at the transition state. Meanwhile, the dependence of shear force on loop size can be calculated from the thermodynamic relation
\begin{equation}\label{force}f(r_0) = -k_B T \left. \frac{\partial \log P(r)}{\partial r} \right|_{r_0} \end{equation}
where $P(r)$ is the equilibrium radial distribution function of end-to-end distance $r$ of a polymer. 

To obtain $P(r)$, we considered two continuous polymer models: the WLC model and the linear subelastic chain (LSEC) model. The WLC model is the canonical elastic DNA model with a quadratic dependence of deformation energy on bending angle. In comparison, the LSEC model assumes a linear relationship between them, and has been proposed as a phenomenological DNA model in the strong bending regime\cite{wiggins2006high,wiggins2006generalized}. The parameters of both models are strongly constrained by the persistence length of ${\sim}$50 nm in the long limit (FIG. ~\ref{sfig2}). When constrained in this fashion, the LSEC model predicts high-curvature conformations more frequently than the WLC model\cite{wiggins2006high,wiggins2006generalized}. We performed the biased Monte Carlo (MC) simulation to calculate the shear force as a function of loop size (see Methods). The LSEC model produces a significantly weaker shear force and a more moderate length-dependence than the WLC model (FIG. ~\ref{fig2}(b)). We note that the calculated shear force depends only weakly on $r_0$ near the value chosen for our analysis (FIG. ~\ref{sfig3}).

\begin{figure}[htp!]
\includegraphics[width=\columnwidth]{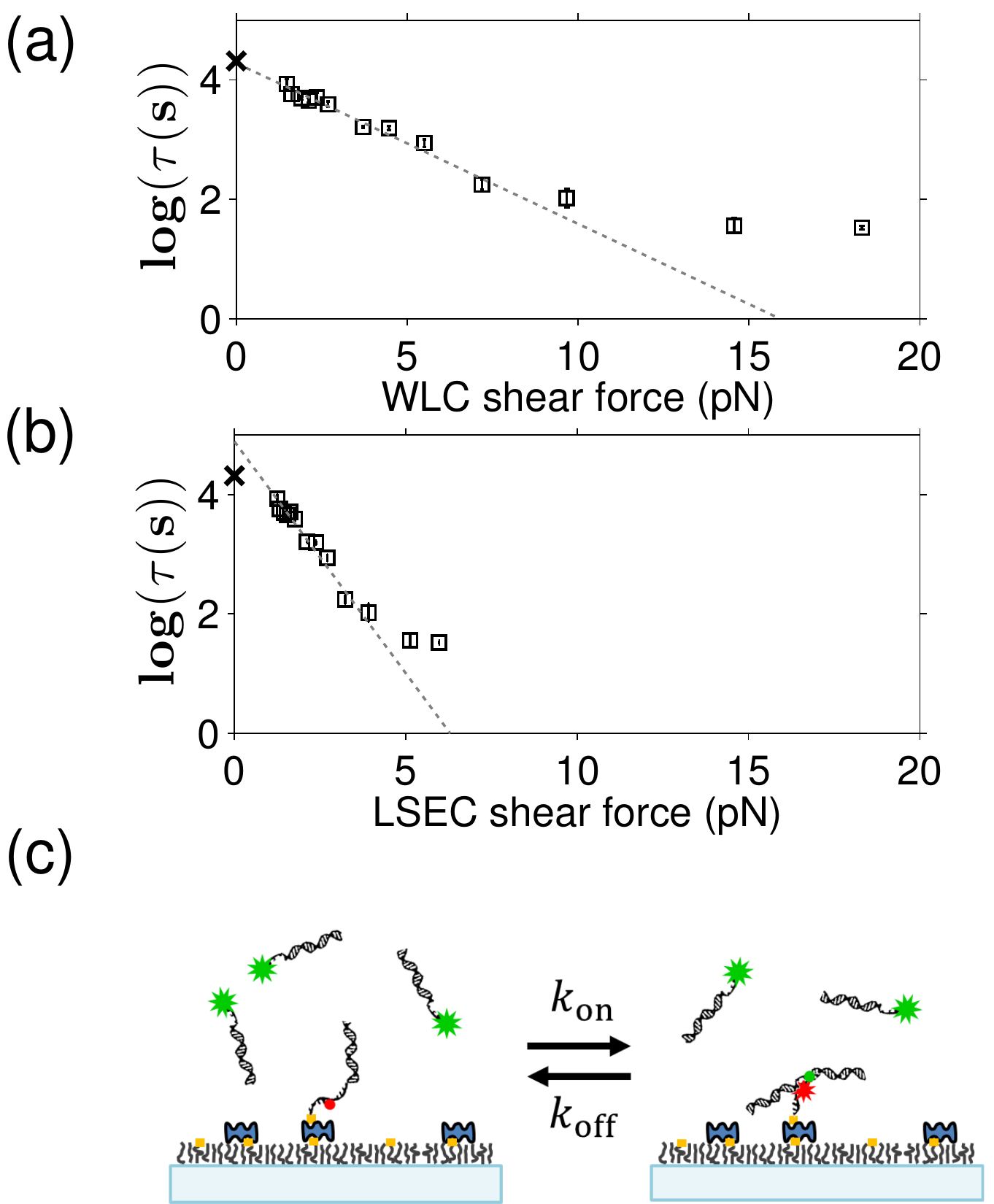}
\caption{\label{fig3}{\bf The relationship between linker lifetime and shear force.} The natural logarithm of the lifetime measured in 50 mM NaCl is plotted as a function of shear force calculated from the WLC (a) and the LSEC model (b). Linear regression yields the zero-force lifetime ($\tau(0)$) and the separation distance ($\Delta r_0$) for duplex dissociation. Only the lifetimes for DNA loops larger than 60 bp are included in the regression. Data for loop size less than 60 bp are excluded from the linear regression based on the RMSE analysis (see Supplementary FIG. ~\ref{sfig4}). (c) The zero-force lifetime ($\tau(0)$) measurement from dissociation kinetics of a linear dimer. In this experiment, the linker formed between the sticky ends of the DNA molecules does not experience a shear force, and therefore, the dissociation lifetime corresponds to $\tau(0)$. DNA molecules are composed of an 18-bp duplex and a 13-mer single-stranded overhang, and are identical to the end-segments of the DNA molecule as depicted in FIG. ~\ref{fig1}(b). The zero-force lifetimes (marked `$\times$') averaged from four measurements are plotted in (a) and (b) for comparison with the two models. The error bar for this data point is smaller than the symbol.}
\end{figure}

We plotted the logarithm of the measured lifetime vs. the calculated forces, which is expected to be a straight line according to Eq. ˜\ref{bell}. As shown in FIG. ~\ref{fig3}, the overall relationship follows a straight line between 60 and 200 bp, but deviates from it at smaller loop sizes (also see the root mean squared error (RMSE) analysis in FIG. ~\ref{sfig4}). This deviation, which corresponds to the inflection point in FIG. ~\ref{fig2}(a), indicates a softening transition of the loop where the actual force becomes weaker than the force predicted by each model. The relationship in the linear regime can be fitted with Eq. ˜\ref{eqb} to obtain the negative slope ($\Delta r_0$) and the y-intercept ($\tau(0)$), both of which are related to the dissociation kinetics of the linker duplex. Since the WLC and LSEC models predict markedly different $\Delta r_0$ (1.10 $\pm$ 0.14 nm vs. 3.18 $\pm$ 0.48 nm) and $\tau(0)$ (72.24 $\pm$ 10.28 sec vs. 132.83 $\pm$ 6.20 sec), we can compare these fitting parameters with experimental values to identify the correct model before the softening transition.

The linker lifetime with zero shear force, $\tau(0)$, can be measured using the same linker without the loop. For this experiment, we prepared two separate DNA molecules identical to the end-segments of the DNA used in the unlooping assay so that they can form the same linker without the shear force (FIG. ~\ref{sfig5} and Supplementary Information). We immobilized the Cy5 DNA on the surface and introduced the Cy3 DNA at ${\sim}$20 nM concentration (FIG. ~\ref{fig3}(b)). Linker formation and separation resulted in two-state fluctuation in Cy5 intensity due to FRET (FIG.~\ref{sfig6}). Linker separation could be well-described by first-order kinetics, from which the lifetime was extracted. We find that the measured $\tau(0)$ (marked `$\times$' in FIG. ~\ref{fig3}(a,b)) agrees well with the WLC model prediction, but not with LSEC.

On the other hand, $\Delta r_0$ was previously measured to be 1 \AA\, per base pair by pulling short DNA duplexes at opposite $5'$-ends\cite{strunz1999dynamic}. In our stretched linker duplex, the total number of complementary base pairs is 13, but the largest number of consecutive base pairs is 9 due to Cy5 in the backbone. Therefore, $\Delta r_0$ can be estimated to be in the range of 0.9 nm to 1.3 nm, which includes the prediction of the WLC model but not the LSEC model. Since both parameters $\Delta r_0$ and $\tau(0)$ are compatible with the WLC model, but not with the LSEC model, we conclude that the free energy of dsDNA loop as small as 60 bp is better described by the WLC model.

\begin{figure*}[htp]
\includegraphics[width=\textwidth]{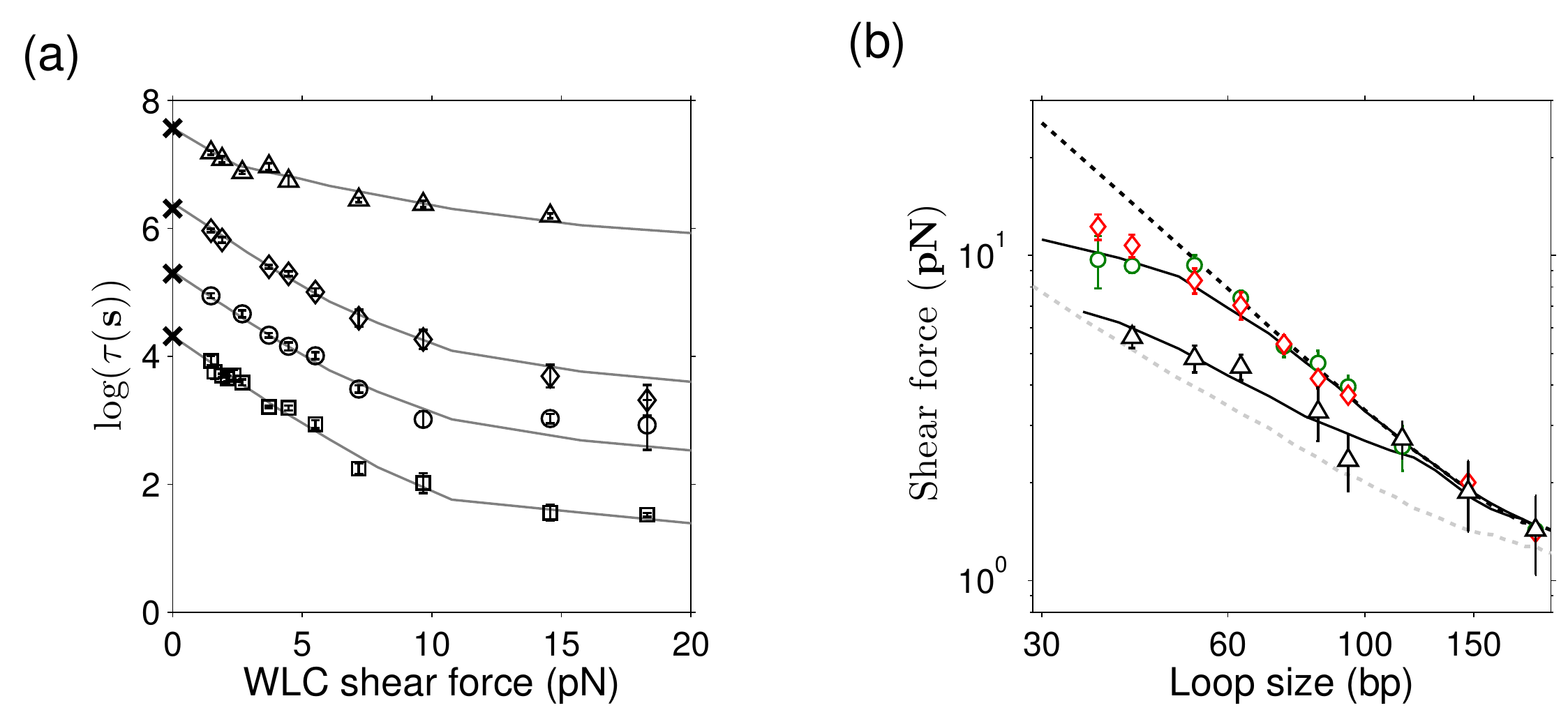}
\caption{\label{fig4}{\bf The effect of sodium and magnesium salt on strong dsDNA bending.} (a) The logarithm of the measured loop lifetime is plotted as a function of the predicted WLC shear force in different [Na\textsuperscript{+}]s (squares: 50 mM, circles: 100 mM, diamonds: 200 mM) and in 5 mM [Mg\textsuperscript{2+}] (triangles). In sodium buffers, the softening transition appears at ${\sim}$8 pN, which corresponds to a loop size of 60 bp whereas in 5 mM [Mg\textsuperscript{2+}], it is noticeable at ${\sim}$3 pN, which corresponds to a loop size of 100 bp. For the WLC model, we assumed a constant persistence length of 50 nm\cite{baumann1997ionic,wenner2002salt}. The lifetimes at zero force (`$\times$' symbols) were measured from the dimer dissociation experiments. For each salt condition, 3 separate measurements were performed. Linear fitting of data points in the elastic regime yields almost identical negative slopes independent of [Na\textsuperscript{+}] concentration. We optimized $h$ and $b$ of the KWLC model to fit the softening transition while fixing $\tau(0)$ and $\Delta r_0$ obtained from the elastic regime. The gray curves are generated with $h = 22 k_B T$ and $b = 0.3$ for 50-200 mM [Na\textsuperscript{+}], and $h = 17 k_B T$ and $b = 0.7$ for 5 mM [Mg\textsuperscript{2+}]. See FIG.~\ref{sfig8} for a zoom-in view. (b) Shear force extracted from the unlooping experiment in 100 mM [Na\textsuperscript{+}] (green circles), 200 mM [Na\textsuperscript{+}] (red diamonds) and 5 mM [Mg\textsuperscript{2+}] (black triangles) are compared with predictions from three DNA models (the axis is in log-log scale). The black curves represent the shear forces calculated from the KWLC model (with parameters listed in (a)). Also shown are the forces calculated from the WLC model (dotted black curve) and the LSEC model (dotted gray curve). The KWLC model with the lower $h$ is similar to the LSEC model while the KWLC model with the higher $h$ is similar to WLC. Below a certain loop size, the KWLC model predicts a smaller shear force than the WLC model because a kink relieves some of the bending stress.}
\end{figure*}

To confirm that our conclusion is not affected by duplex dissociation kinetics, we conducted the unlooping assay at different [Na\textsuperscript{+}] concentrations. In the range between 50 and 200 mM [Na\textsuperscript{+}], $\tau(0)$ is expected to increase with [Na\textsuperscript{+}]\cite{braunlin1991proton,dupuis2013single} whereas the persistence length of dsDNA should not depend on [Na\textsuperscript{+}]. As expected, the linker lifetime $\tau(0)$ was significantly prolonged at higher salt concentrations (FIG. ~\ref{fig4}(a)). Despite changes in loop lifetimes as a function of [Na\textsuperscript{+}], all curves exhibit a softening transition near 60 bp, and all $\tau(0)$ values (marked `$\times$' in FIG.~\ref{fig4}(a)) overlap nicely with the values extrapolated by the WLC model. The observed relationships at different [Na\textsuperscript{+}] also collapsed to the same line when normalized by $\tau(0)$ (Supplementary FIG. ~\ref{sfig7}). This result further supports our conclusion that the WLC model correctly describes the free energy of dsDNA bending prior to the softening transition.

We also investigated how magnesium affects strong bending of dsDNA. Magnesium is essential for the activity of the ligase in the cyclization assay, and the restriction enzyme in DNA minicircle digestion. Therefore, almost all enzyme-based experiments on strong dsDNA bending have been performed in the presence of magnesium at relatively high concentrations (5-10 mM). Interestingly, we found that in the presence of 5 mM [Mg\textsuperscript{2+}], the softening transition of dsDNA occurs near 100 bp (FIG.~\ref{fig4}(a) and Supplementary FIG. ~\ref{sfig8}). This result indicates that magnesium can dramatically increase the apparent flexibility of dsDNA in the strong bending regime.

\section{\label{dis}Discussion}
Using a FRET-based unlooping assay, we probed the energetics of dsDNA bending in the strong bending regime. We measured the loop lifetime as a function of loop size. In standard Na\textsuperscript{+} concentrations between 50 and 200 mM, the observed relationship in the range between 60 and 200 bp was consistent with the WLC model. Below 60 bp, we observed that dsDNA loses elastic rigidity, which leads to a weaker dependence of the shear force on the loop size. The critical loop size where softening occurs corresponds to a maximum bending angle of $7^{\circ}$/bp in a teardrop shape. In the presence of 5 mM [Mg\textsuperscript{2+}], the critical loop size increased to 100 bp, corresponding to $4^{\circ}$/bp. This result suggests that in cyclization experiments that typically use 10 mM [Mg\textsuperscript{2+}], subelastic bending can enhance the looping probability of dsDNA shorter than 100 bp. 

The interpretation of our results relies on the Bell relationship between duplex lifetime and stretching force\cite{bell1978models}. In general, a bond can dissociate through several different pathways\cite{hinczewski2013mechanical}, which may give rise to a nontrivial relationship between bond lifetime and the applied force\cite{evans2004mechanical}. However, our assumption of the Bell model is justified by previous experimental studies\cite{strunz1999dynamic,comstock2011ultrahigh}. Notably, a DNA duplex pulled at the opposite $5'$-ends by AFM, in the same shear geometry as in our DNA loop, exhibited strand separation kinetics consistent with a single energy barrier along the mechanical separation path. Also, the Chemla group recently demonstrated that DNA duplex dissociation under a constant tensile force follows the Bell relationship by combining fluorescence with optical tweezers\cite{comstock2011ultrahigh}. In that study, the relationship between $\Delta r_0$ and duplex length ($L$) was extracted to be $\Delta r_0=0.096 \times L$ (nm), and has been more precisely determined as $\Delta r_0=0.256 \times (L-6)$ (nm) (personal communication with Dr. Chemla). Either estimation puts $\Delta r_0$ to be in the range consistent with the WLC model but not with the LSEC model.

The breakdown of continuous models below the critical loop size is likely due to structural transition in the dsDNA helix such as kink formation\cite{crick1975kinky,lankavs2006kinking} that renders DNA softer. For free DNA, kinks are rare, transient deformations only occurring at a rate of $10^{-4}-10^{-5}$\cite{zeida2012breathing,frank2014fluctuations}, but they can become significant in sharply bent DNA\cite{lankavs2006kinking,lee2012enhanced,fields2013euler}. We can use the apparent critical loop size to set the lower limit on the energy barrier for kink formation. To account for the effect of kinking on loop stability, we consider the kinkable worm-like chain (KWLC) model\cite{yan2004localized,wiggins2005exact,vologodskii2013strong} cast in a simple functional form proposed by Vologodskii and Frank-Kamenetskii\cite{vologodskii2013strong}. In this model, the dinucleotide bending energy ($E$) is given by 
\begin{equation}\label{kink}E = \min \left(\frac{1}{2} k \theta^2 , h + (\theta - b)^6 \right) \end{equation}
where $k$ is the bending rigidity identical to that of the WLC model, $h$ is the energy barrier of kinking, and $b$ specifies the range of bending angles at the kink. We varied $h$ while fixing $b$ in our simulation to find $h$ that is most compatible with the observed critical length of 60 bp. The parameter b was chosen to be 0.3 which allows kink angles up to $90^\circ$\cite{vologodskii2013strong} based on other calculations and molecular dynamics simulations\cite{crick1975kinky,lankavs2006kinking,mitchell2011atomistic}. As shown in FIG. ~\ref{fig4}, $h=22 k_B T$ and $b=0.3$ can produce a transition in the shear force below 60 bp, which is consistent with our observation. Using this $h$ value, we can also calculate the free energy of kink formation ($\Delta G_k$) (more details in the Supplementary Information) to be $\Delta G_k \approx 18 k_B T$, which is similar to the upper limits of previous estimations\cite{zheng2009theoretical,fields2013euler}. In comparison to $h=22 k_B T$ and $b=0.3$ in the KWLC model, the lifetime vs. loop size relationship taken at 5 mM [Mg\textsuperscript{2+}] yields $h=17k_B T$ and $b =0.7$. These parameters correspond to a lower free energy of kink formation of $\Delta G_k = 12 k_B T$ and larger kink angles up to $110^\circ$. 

Using the parameters, h and b, constrained by our data, we can also determine the probability of kink formation in a DNA minicircle as a function of loop size. We performed a restrained MC simulation of DNA minicircles of various sizes (see Methods) to measure the frequency of large angle deflections in thermal equilibrium. In our simulation, we only consider the effect of bending stress on kink formation. As shown in FIG.~\ref{sfig10}, in the absence of magnesium, kink formation is negligible even in 60-bp loops due to a high energy barrier. In the presence of 5 mM [Mg\textsuperscript{2+}], however, the kinking probability increases sharply with decreasing loop size, approaching unity at 70 bp while remains insignificant for DNA over 100 bp. This simulation result agrees well with a previous minicircle digestion study that detected kinks in 60-bp minicircles due to bending stress alone\cite{du2008kinking}.

\begin{figure}[h]
\includegraphics[width=\columnwidth]{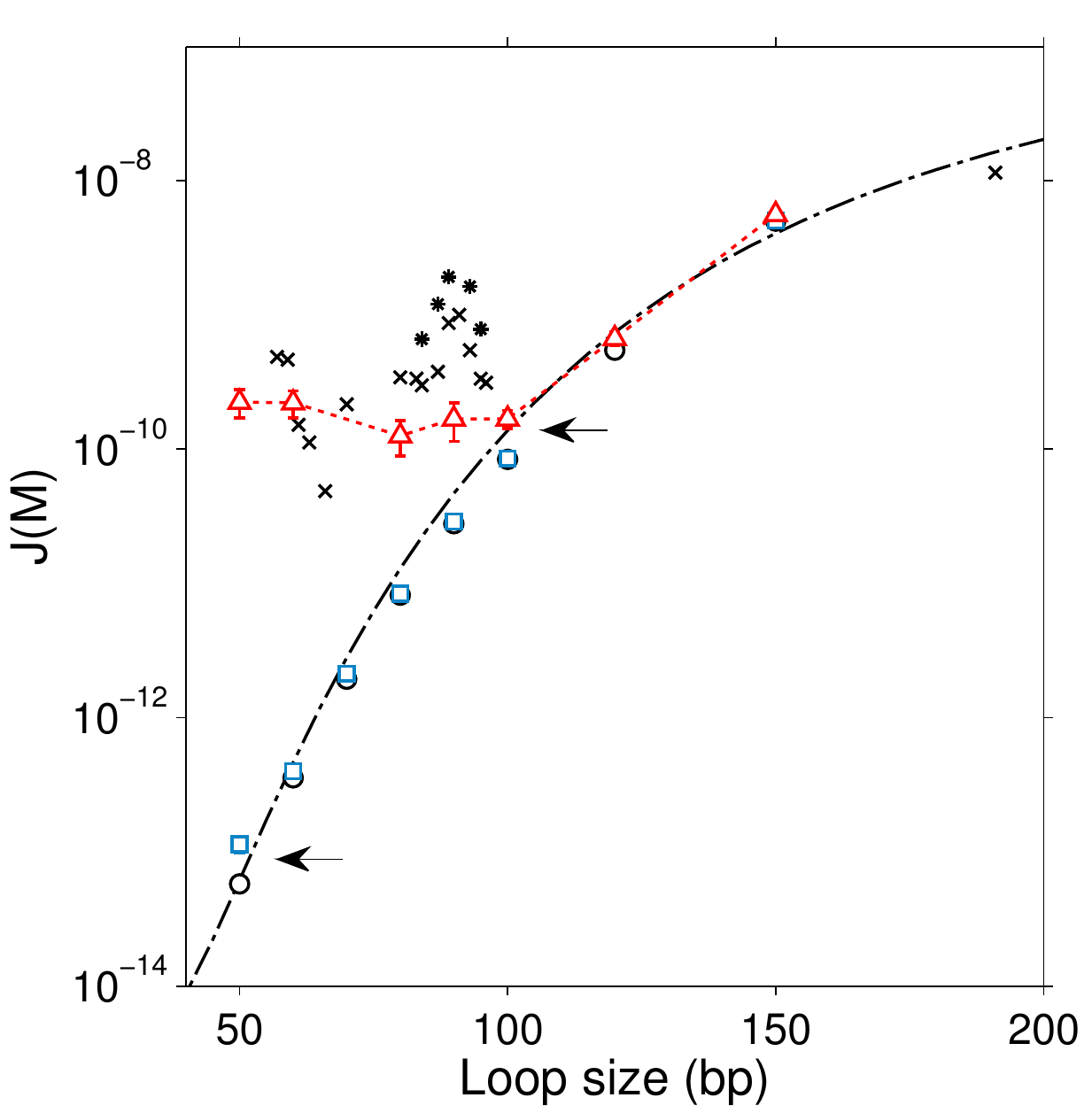}
\caption{\label{fig5}{\bf J factor comparison}. The J factor was computed using a fixed end-to-end distance of 5 nm without end-to-end angular or torsional constraints. We used the weighted histogram analysis method to calculate J factors predicted by the WLC model (black circles), the KWLC model with $h=17 k_B T$ and $b = 0.7$ (red triangles), and with $h = 22 k_B T$ and $b = 0.3$ (blue squares). The dash-dotted line is the WLC-prediction according to the theoretical approximation\cite{douarche2005protein}. For comparison, the J factors from Vafabakhsh and Ha\cite{vafabakhsh2012extreme} are also shown. Symbols marked `$\times$' indicate J factors measured from surface-immobilized DNA in 1 M [Na\textsuperscript{+}] and symbols marked `$\ast$' from vesicle-encapsulated DNA at 10 mM [Mg\textsuperscript{2+}]. For consistency with our calculation, we shifted the original data\cite{vafabakhsh2012extreme} by 10 bp to the left to account for the linker length. The arrows indicate where the KWLC model deviates from the WLC model. We performed 4 simulations to generate the SEM error bars for each J factor.}
\end{figure}

Our results suggest that magnesium can promote subelastic bending above a critical bending angle of $4^{\circ}$/bp by stabilizing large-angle deformations. This interpretation is similar to the conclusion of a recent study with DNA vises\cite{fields2013euler}. Therefore, we considered whether magnesium-facilitated softening could explain high J factors reported previously\cite{vafabakhsh2012extreme}. We thus calculated the J factor as a function of length using the KWLC model with $h = 17 k_B T$ and $b=0.7$ constrained by the data taken at 5 mM [Mg\textsuperscript{2+}]. As shown in FIG.~\ref{fig5}, while the KWLC model produces J factors similar to the WLC model prediction above 100 bp as previously demonstrated\cite{du2005cyclization}, it also predicts substantially higher J factors for DNA below 100 bp, matching the J factors determined from the single-molecule FRET cyclization study\cite{vafabakhsh2012extreme} within a factor of 10. The agreement between our KWLC model and the result of the FRET study may be closer if the difference in the buffer condition (5 mM vs. 10 mM [Mg\textsuperscript{2+}]) and the uncertainty associated with J factor determination is accounted for (Supplementary Information). In the absence of magnesium, however, our KWLC model predicts that the WLC model will be valid at least down to 55 bp (blue squares, FIG.~\ref{fig5}). This may explain why some studies lacking magnesium did not observe enhanced dsDNA flexibility at short length scales\cite{schopflin2012probing,shi2013structural}.

Our unlooping assay enables investigation of strong dsDNA bending in buffer conditions not compatible with the ligation-based cyclization, the FRET-based cyclization, and the AFM assay. In the ligation assay, magnesium must be present at high concentrations for ligase activity. For AFM, magnesium is necessary to bind DNA to the surface\cite{mantelli2011conformational}. In the FRET-based cyclization assay, high magnesium or sodium concentration is necessary to produce a statistically significant number of looping events. In this study, we demonstrated that effects of small amounts of monovalent and divalent ions on the elastic limit of dsDNA can be studied separately. Moreover, the unlooping assay is more well-suited to the study of kink formation than the cyclization assay because the probability of kink formation increases with bending stress. Our unlooping assay is similar in some ways to previous methods employing small DNA loops\cite{shroff2008optical,wang2013critical,fields2013euler}. In these studies, electrophoretic mobility or intramolecular FRET of these loops was measured to investigate kinking. Our approach differs from theirs in two ways. First, we measure kinetic decay of the looped state instead of equilibrium distribution between alternative conformations in the looped state. Second, we do not need to include stretching or twisting energy in the Hamiltonian for single-stranded parts or twisted dsDNA. Therefore, our method allows a more direct link between the measurable quantities and dsDNA bending rigidity and holds great promise for studying the effect of sequence, salt, and temperature on strong dsDNA bending.

\section*{\label{me}Methods}
\paragraph{Materials}
The DNA molecules used in this unlooping assay have a double-stranded part with variable length from 37 bp to 189 bp and 13-nucleotide (nt) long single-stranded complementary overhangs (sticky ends). One overhang contains Cy3, and the other contains Cy5 and biotin (FIG. ~\ref{fig1}b). The sequences of these overhangs are ATAG/iCy5/GAATTTACC, where /iCy5/ represents the internally labeled Cy5, and GGTAAATTC\underline{A}CTAT with the underlined `A' inserted as a spacer opposite to iCy5 to increase the likelihood of base pairing around iCy5 that interrupts the backbone. All DNA molecules are derived from a master sequence that is ${\sim}$50 \% in GC content and does not have curvature-inducing patterns such as GGGCCC or A-tracts. The master sequence was constructed by annealing the ends of two 113-nt long single-stranded DNAs over a 16-nt region and extending their $3'$-ends using DNA polymerase. The 210-bp master DNA was purified by gel electrophoresis, and PCR-amplified with dangling-end primers to generate DNAs with common terminating sequences. The annealing location of one of the primers was varied to generate DNAs with different lengths. These PCR products were used as templates in another round of PCR to incorporate fluorescent labels and a biotin as previously described\cite{le2013measuring}.  Strands were exchanged between these PCR products to obtain the final DNA constructs for our experiment. Detailed sequences can be found in the Supplementary Information.

\paragraph{Single-molecule unlooping assay}
The DNA molecules were immobilized on a PEG-coated glass surface through NeutrAvidin-biotin interaction. The immobilized molecules were excited by the evanescent wave of a 532-nm laser (NT66-968, B\&W Tek, Newark, DE) totally internally reflected through a high NA objective (UApo N 100$\times$/1.49, Olympus). The power of the 532 nm laser was ${\sim}5\mu$W when measured after the microscope objective before reaching the critical angle of incidence.  For a split view of Cy3 and Cy5 images, the fluorescence image was split into the Cy3 and Cy5 channels outside the microscope and relayed onto an EMCCD (DU-897ECS0-\# BV, Andor). A lab-written C program was used to view and save live images from the CCD. The raw image data were processed by MATLAB to generate single-molecule time traces of Cy3 and Cy5 intensities.   
In the loop breakage assay, immobilized DNA molecules were first incubated in 2 M NaCl buffer for up to an hour to generate looped molecules. We then introduced the imaging buffer (5 mM PCA, 100 mM PCD, 1 mM Trolox) that contains 2 M NaCl to start image acquisition. After 20 seconds, new imaging buffer with 50-200 mM NaCl was perfused into the imaging channel at a flow rate of 75 $\mu$L/min, which corresponds to ${\sim}$1 cm/s in flow velocity through the channel. The typical dimension of the channel cross-section is 0.075 mm $\times$ 2.0 mm. We recorded the times it takes for molecules to unloop from single-molecule time traces, built the survival time histogram, and fitted it with a single exponential function to extract the linker lifetime. 

\paragraph{Force calculation for different DNA models}
To calculate the shear force (Eq.˜\ref{force}), we used umbrella sampling to generate the radial probability distribution (P(r)). dsDNA was treated as a chain of rigid monomers, and bending energy was assigned to each angle between adjacent monomers. Thus, the Hamiltonian was the sum of the total bending energy of the polymer from all monomer steps $\sum_{i=1}^{N-1} \frac{k}{2} \theta_{i,i+1}^{2}$ for WLC and $\sum_{i=1}^{N-1} B |\theta_{i,i+1}|$ for LSEC where $\theta_{i,i+1}$ is the angle between the i-th monomer and the i+1-th monomer) and the harmonic potential ($\frac{1}{2}K (r-r_0)^2$) with stiffness $K$ that restrains the end-to-end distance near $r_0$. For the WLC model, each base pair was treated as a monomer, similar to the dinucleotide model. For the LSEC model, we tried 7-bp long monomers as published\cite{wiggins2006generalized,wiggins2006high}. The bending rigidity constants were chosen so that both models predict a persistence length of ${\sim}$50 nm in the long limit (see Supplementary Information). We also considered the KWLC model\cite{vologodskii2013strong} that allows for kink formation at large bending angles. The bending energy for the (i,i+1)-th dinucleotide step is $\min \left(\frac{1}{2} k \theta_{i,i+1}^2 , h + (\theta_{i,i+1} - b)^6 \right)$. In this formula, $k$ is the bending rigidity which is the same as in the WLC model, and $h$ is the energy barrier for kinking\cite{vologodskii2013strong}. $b$ was fixed to 0.3 radians (if not mentioned otherwise) to allow the kinks to adopt bending angles up to 90$\,^{\circ}$. 

Except for the bias potential for umbrella sampling, we did not apply constraints on relative bending or torsional angles between the two ends because flexible gaps at the ends of the linker effectively relax bending and torsional stress. The lack of angular constraints in the loop geometry of our DNA construct is supported by the observation that the J-factor of DNA with gaps does not oscillate with the helical phase of DNA\cite{du_gapped_2005}, in contrast to intact DNA circles\cite{peters2010dna,vologodskaia_contribution_2002}. 

In principle, the force can be obtained from the derivative of the unbiased radial probability distribution at $r_0$ according to Eq.˜\ref{force}. Because short distances are rarely populated, we used umbrella sampling where a biasing harmonic potential of stiffness $K$ is applied near $r_0$ to obtain a sufficient number of looped conformations. The spring constant $K$ for the biasing potential in the case of the WLC model and the LSEC model was set to 8 pN $\cdot$ nm/(1 bp)\textsuperscript{2} and 400 pN $\cdot$ nm/(7 bp)\textsuperscript{2}, respectively. The biased force ($f^b$) is then given by\cite{kastner2005bridging}
\begin{equation}\label{biased}f^b (r)=f^u (r)+K(r-r_0). \end{equation}
Thus, the unbiased force ($f^u$) is equal to $f^b$ if evaluated at $r_0$, which enables us to use Eq.˜\ref{biased} to calculate $f^u$ directly from a biased radial probability distribution. Since derivatives are sensitive to statistical noise, we instead used an approximation that contains averaging\cite{hwang2007calculation}
\begin{equation}f(r_0)=-\frac{k_B T}{\text{var}(\delta r)}<\delta r>,\end{equation}
where $\delta r$ is the deviation of the end-to-end distance from $r_0$. $<\delta r>$ and $\text{var}(\delta r)$ are the mean and the variance of these deviations, respectively. Pivot moves were used to sample the conformational space of the chain, and Metropolis criterion was applied to accept conformations consistent with the Boltzmann distribution. The chain was equilibrated for $10^5$ MC steps starting from the minimum energy conformation, and approximately $5\times10^6$ conformations after equilibration were used to obtain $P(r)$. The calculated force for a specific loop size did not depend on the value of $K$. For the WLC and the LSEC models with monotonically increasing bending energy, the calculated force varied little between simulations. For the KWLC model with a discontinuous slope, the calculated force for small loop sizes was more variable and, therefore, we increased the number of simulations until the SEM was smaller than 8\% of the mean.

\paragraph{Analysis of linker lifetime vs. force}To analyze the linker lifetime vs. shear force, we performed linear regression with the `robustfit' function (MATLAB). We also examined how the goodness of fit changes with the range of fitting using the standard regression error or RMSE (root mean squared error) as an indicator. As shown in FIG. ~\ref{sfig4}, the RMSEs for both WLC and LSEC models increase significantly when points below ${\sim}$60 bp were included. This analysis indicates that Eq.˜\ref{bell} does not hold below this length because the calculated forces are overestimated compared to the actual forces exerted on the linker. Therefore, we did not include these points when extracting the fitting parameters, $\tau (0)$ and $\Delta r_0$ for WLC and LSEC models.

\paragraph{J factor calculation}
The J factor is calculated by Weighted Histogram Analysis Method (WHAM)\cite{kumar1992weighted,becker2010radial}. A number of umbrella sampling simulations were carried out, each having its own restraint energy $U_j(r_k)$ where $j$ is the simulation index, and $k$ is the bin index. In the $j$-th simulation, one obtains the number of counts $n_{j,k}$ in the $k$-th bin with the total counts equal to $N_j=\sum_k n_{j,k}$. Using the bias factor in each bin $c_{j,k}=\exp (-U_j(r_k))$, we can obtain the radial probability density of the unrestrained chain ($p_k^0$)
\begin{subequations}
 \label{wham}
 \begin{eqnarray}
 p_k^0 = \frac{\sum_j n_{j,k}}{\sum_j f_{k,j}N_j}, \label{eq1}\\
 f_{k,j}= \frac{c_{j,k}}{\sum_k c_{j,k}p_k^0}. \label{eq2}
 \end{eqnarray}
\end{subequations}
These equations were solved iteratively by updating the equations until $p_k^0$ converges. We adjusted the spring constant and restraint coordinates so that there is significant overlap between adjacent histograms. Typically, each individual histogram was built from $10^6$ chains. The J factor was obtained by normalizing $p_k^0$, dividing it by $4\pi r^2$, and converting it to molar units.  

\paragraph{Minicircle simulations}
The MC simulation for a DNA minicircle was implemented as previously described\cite{zheng2009theoretical}. We applied the KWLC bending energy to each link and calculated the total bending energy of the minicircle. Random conformations generated by crankshaft rotations were selected based on the Metropolis criteria. In one course of simulation, $15 \times 10^6$ conformations were  typically collected. To enhance the sampling efficiency, we randomly picked angles for the crankshaft rotation from two uniform distributions across two intervals, [$-90^\circ$, $90^\circ$] and [$-10^\circ$, $10^\circ$]. For each accepted conformation, all the dinucleotide angles were recorded to determine if the minicircle has kinks. A kink was assigned if the bending angle exceeds the critical kink angle defined as the intercept of the two energy terms in Eq.~\ref{kink}. For each loop size, we calculated the kinking probability, which is the fraction of accepted conformations with at least one kink.   
 
\acknowledgements
We thank the lab members for critical discussions and reading of the manuscript. H.D.K and co-workers are funded by the Georgia Institute of Technology start-up funds, the Burroughs Welcome Fund Career Award at the Scientific Interface, and Physics of Living Systems student research network grant from the National Science Foundation. T.T.L acknowledges the financial support from the GANN and COS Molecular Biophysics Students funding. 
    
\section*{Authors contributions}
T.T.L. and H.D.K. designed the study. T.T.L prepared the sample and collected data. T.T.L and H.D.K analyzed the data and wrote the paper.

\section*{Competing interests}
The authors declare no competing financial interests. 

\clearpage

\beginsupplement
\onecolumngrid
\section*{Supplementary Information}
\subsection*{DNA sequences (from $5'\rightarrow 3'$)}
\subsubsection{Master 210 bp DNA}

\noindent\begin{minipage}{\textwidth}
\noindent\rule{\textwidth}{.5pt}
\DNA! gtgccagcaacagatagcctatccatagactattacctacaagcccaatagcgtacgggatcatccccgccagttacgtctgccacccttcttaacgacacgtgaagggacgaaccgcatacttacgatcaggcatagatcttacaccgtagcaggtagtgccaggcatcgtgttcgtaaccttacttcaaccattcgagctcgttgttg !
\end{minipage}

\subsubsection{189 bp} 
\noindent\begin{minipage}{\textwidth}
\noindent\rule{\textwidth}{.5pt}
\DNA! gtgccagcaacagatagcctatccatagactattacctacaagcccaatagcgtacgggatcatccccgccagttacgtctgccacccttcttaacgacacgtgaagggacgaaccgcatacttacgatcaggcatagatcttacaccgtagcaggtagtgccaggcatcgcattcgagctcgttgttg !
\end{minipage}

\subsubsection{168 bp} 
\noindent\begin{minipage}{\textwidth}
\noindent\rule{\textwidth}{.5pt}
\DNA! gtgccagcaacagatagcctatccatagactattacctacaagcccaatagcgtacgggatcatccccgccagttacgtctgccacccttcttaacgacacgtgaagggacgaaccgcatacttacgatcaggcatagatcttacaccgtcattcgagctcgttgttg !
\end{minipage}

\subsubsection{147 bp} 
\noindent\begin{minipage}{\textwidth}
\noindent\rule{\textwidth}{.5pt}
\DNA! gtgccagcaacagatagcctatccatagactattacctacaagcccaatagcgtacgggatcatccccgccagttacgtctgccacccttcttaacgacacgtgaagggacgaaccgcatacttacgatcattcgagctcgttgttg !
\end{minipage}

\subsubsection{136 bp} 
\noindent\begin{minipage}{\textwidth}
\noindent\rule{\textwidth}{.5pt}
\DNA! gtgccagcaacagatagcctatccatagactattacctacaagcccaatagcgtacgggatcatccccgccagttacgtctgccacccttcttaacgacacgtgaagggacgaaccgccattcgagctcgttgttg !
\end{minipage}

\subsubsection{126 bp} 
\noindent\begin{minipage}{\textwidth}
\noindent\rule{\textwidth}{.5pt}
\DNA! gtgccagcaacagatagcctatccatagactattacctacaagcccaatagcgtacgggatcatccccgccagttacgtctgccacccttcttaacgacacgtgaaggcattcgagctcgttgttg !
\end{minipage}

\subsubsection{115 bp} 
\noindent\begin{minipage}{\textwidth}
\noindent\rule{\textwidth}{.5pt}
\DNA! gtgccagcaacagatagcctatccatagactattacctacaagcccaatagcgtacgggatcatccccgccagttacgtctgccacccttcttaacgcattcgagctcgttgttg !
\end{minipage}

\subsubsection{94 bp} 
\noindent\begin{minipage}{\textwidth}
\noindent\rule{\textwidth}{.5pt}
\DNA! gtgccagcaacagatagcctatccatagactattacctacaagcccaatagcgtacgggatcatccccgccagttacattcgagctcgttgttg !
\end{minipage}

\subsubsection{84 bp} 
\noindent\begin{minipage}{\textwidth}
\noindent\rule{\textwidth}{.5pt}
\DNA! gtgccagcaacagatagcctatccatagactattacctacaagcccaatagcgtacgggatcatcccattcgagctcgttgttg !
\end{minipage}

\subsubsection{74 bp} 
\noindent\begin{minipage}{\textwidth}
\noindent\rule{\textwidth}{.5pt}
\DNA! gtgccagcaacagatagcctatccatagactattacctacaagcccaatagcgtaccattcgagctcgttgttg !
\end{minipage}

\subsubsection{63 bp} 
\noindent\begin{minipage}{\textwidth}
\noindent\rule{\textwidth}{.5pt}
\DNA! gtgccagcaacagatagcctatccatagactattacctacaagcccattcgagctcgttgttg !
\end{minipage}

\subsubsection{53 bp} 
\noindent\begin{minipage}{\textwidth}
\noindent\rule{\textwidth}{.5pt}
\DNA! gtgccagcaacagatagcctatccatagactattacattcgagctcgttgttg !
\end{minipage}

\subsubsection{42 bp} 
\noindent\begin{minipage}{\textwidth}
\noindent\rule{\textwidth}{.5pt}
\DNA! gtgccagcaacagatagcctatcccattcgagctcgttgttg !
\end{minipage}

\subsubsection{37 bp} 
\noindent\begin{minipage}{\textwidth}
\noindent\rule{\textwidth}{.5pt}
\DNA! gtgccagcaacagatagcccattcgagctcgttgttg !
\end{minipage}

\subsection*{Preparing partially hybridized DNA molecules for $\tau(0)$ measurement (italic: double-stranded region)}

Cy3-DNA:

$5'$ - Cy3-ggtaaattcactat \textit{caacaacgagctcgaatg} - $3'$

$3'$ - \textit{gttgttgctcgagcttac} - $5'$                 (blocking oligo)

Cy5-DNA:

$5'$ – BiotinTEG - gaaacatag/ iCy5 /gaatttacc \textit{gtgccagcaacagatagc} - $3'$

$3'$ - \textit{cacggtcgttgtctatcg} - $5'$        (blocking oligo)

We mixed equal amounts of the two partially hybridized DNA molecules in annealing buffer (100mM NaCl, 10 mM Tris$\cdot$HCl pH 7.0, 1 mM EDTA) to obtain a final concentration of 5 $\mu$M. The mixture was heated at 95$^\circ$C for 5 minutes, slowly cooled down to room temperature, and loaded on a polyacrylamide gel (19:1 Acryl:Bis, 15\% (w/v) in TBE 1X pH 8.0). Linear dimers were extracted from the gel using an electroelution kit (G-CAPSULE, 786-001, G-Biosciences) after running the gel at 10 V/cm for ${\sim}$1 hour (see Supplementary FIG. ~\ref{sfig5}).

\subsection*{Shear force vs. loop length}
\begin{figure}
\begin{minipage}[c][\textheight]{\textwidth}
\includegraphics[width=12cm]{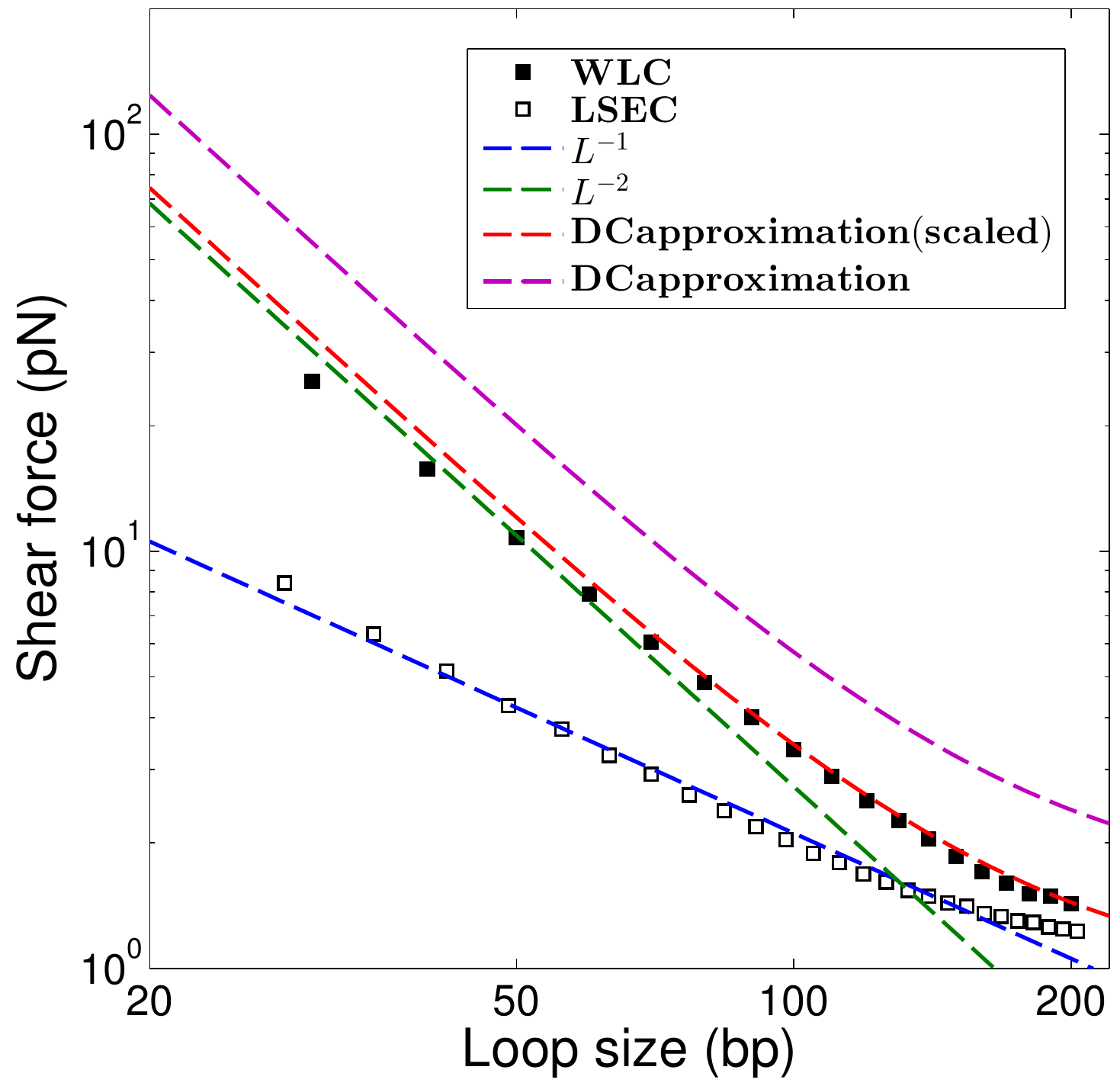}
\caption{\label{sfig1}{\bf The shear force vs. loop size.} The symbols are obtained from the MC simulation. Solid and hollow squares are for WLC and LSEC, respectively. The relationship is plotted on log-log axes to highlight the scaling. For reference, inverse (blue) and inverse-square (green) laws are shown. Douarche and Cocco (DC) approximation with no adjustment (purple) and with scaling by a factor of 0.6 (red) are also shown.}
\end{minipage}
\end{figure}

Here, we derive an approximate relationship between the total bending energy of a circular loop and loop length ($L$). From this relationship, we can obtain the shear force. We assume that the loop takes the shape of a circular arc with the two ends separated by distance $r$. If the bending rigidity of the chain is $k$, the total bending energy of the loop is calculated as
 
\begin{equation}\label{e1} E(r) = \frac{k}{2} \int_{0}^{L}{ds \frac{1}{R(s)^n}} \approx \frac{k}{2} \int_{0}^{L}{ds \frac{1}{ \left(\frac{r+L}{2\pi}\right)^n}} = 2\pi^2 k\frac{L}{ (r+L)^n}, \end{equation}
where $s$ is the distance coordinate along the contour, and $R$ is the radius curvature, which is constant for a circular arc. $n$ is $1$ for the linear subelastic chain model, and $2$ for the WLC model. We make the assumption that $r$ is much smaller than $L$. Differentiating the bending energy with $r$, we can obtain the shear force acting along $r$,
 
\begin{equation}\label{f1} f = -2\pi^2 n k  \frac{L}{(r+L)^{n+1}}. \end{equation}
Thus, at short end-to-end distances, we expect the shear force to scale as $L^{-2}$ for a worm-like chain, and  $L^{-1}$ for a subelastic chain. As shown in FIG. ~\ref{sfig1}, this approximate expression can explain the scaling force vs. length computed from the MC simulation to some degree. However, it overestimates the absolute force values because the dominant loop conformation of a worm-like chain is closer to a teardrop, which is overall less stressed than a circular arc.

A more accurate description of the shear force requires the full probability distribution of end-to-end distances.  An exact analytical expression does not exist in a closed form, and therefore, we use an approximation that best describes the probability distribution at short end-to-end distances in the stiff limit\cite{becker2010radial}. Douarche and Cocco proposed such approximation (DC approximation) that considers both the Boltzmann weight due to the elastic energy of the loop and the fluctuation around the minimum energy conformation\cite{douarche2005protein,allemand2006loops}. The cyclization factor is given by\cite{allemand2006loops}
 
\begin{equation}\label{j1} j(L,r) = \frac{1.66 \times 112.04}{{L_p}^3} \left(\frac{L+2r}{L_p}\right)^{-5} e^{0.246 \frac{L+2r}{L_p}} e^{-14.055\frac{L_p}{L+2r}}. \end{equation}
Multiplying this by $4\pi r^2$ and differentiating,
 
\begin{equation}\label{f2} f = -k_B T \left(\frac{2}{r} - \frac{10}{L+2r} + \frac{0.492}{L_p} +\frac{28.11 L_p}{L^2}\right). \end{equation}
Using $L_p$ = 50 nm  and $r$=5 nm (14.7 bp), we obtain the relationship between the shear force in piconewton and DNA length in units of base pair number ($N_{bp}$).
 
\begin{equation}\label{f3} f[\mathrm{pN}] = -\frac{49849}{{N_{bp}}^2} - 1.68 + \frac{120.5882}{N_{bp} + 29.4}. \end{equation}
This expression with no further adjustment can well describe the scaling of the relationship, but overestimates the force almost by a constant scaling factor. If we multiply the force by 0.6, we find an excellent agreement across the length range of interest. It is not surprising that the  approximation overestimates the absolute force value. When compared with the exact density, the DC probability density is shown to have a steeper slope at short extension\cite{becker2010radial}, which results in slightly higher force values.   

\subsection*{Parameter choice for polymer models}
The length of the monomer and the value of the rigidity constant are chosen so that the known statistical mechanical properties of the polymer in the long limit can be reproduced by simulation. In the case of dsDNA, these parameters can be determined based on the persistence length ($L_p$) of the polymer, which is approximately 50 nm. A linear dsDNA molecule longer than the persistence length can be well described as a worm-like chain, and the mean-square end-to-end distance $\langle R^2 \rangle$ is related to its contour length $L$ as
\begin{equation}
\label{r2}  
\langle R^2 \rangle = 2 L_p L \left[ 1 - \frac{L_p}{L} \left(1 - e^{-L/L_p}\right)\right].
\end{equation}

\begin{figure}
\begin{minipage}[c][\textheight]{\textwidth}
\includegraphics[width=12cm]{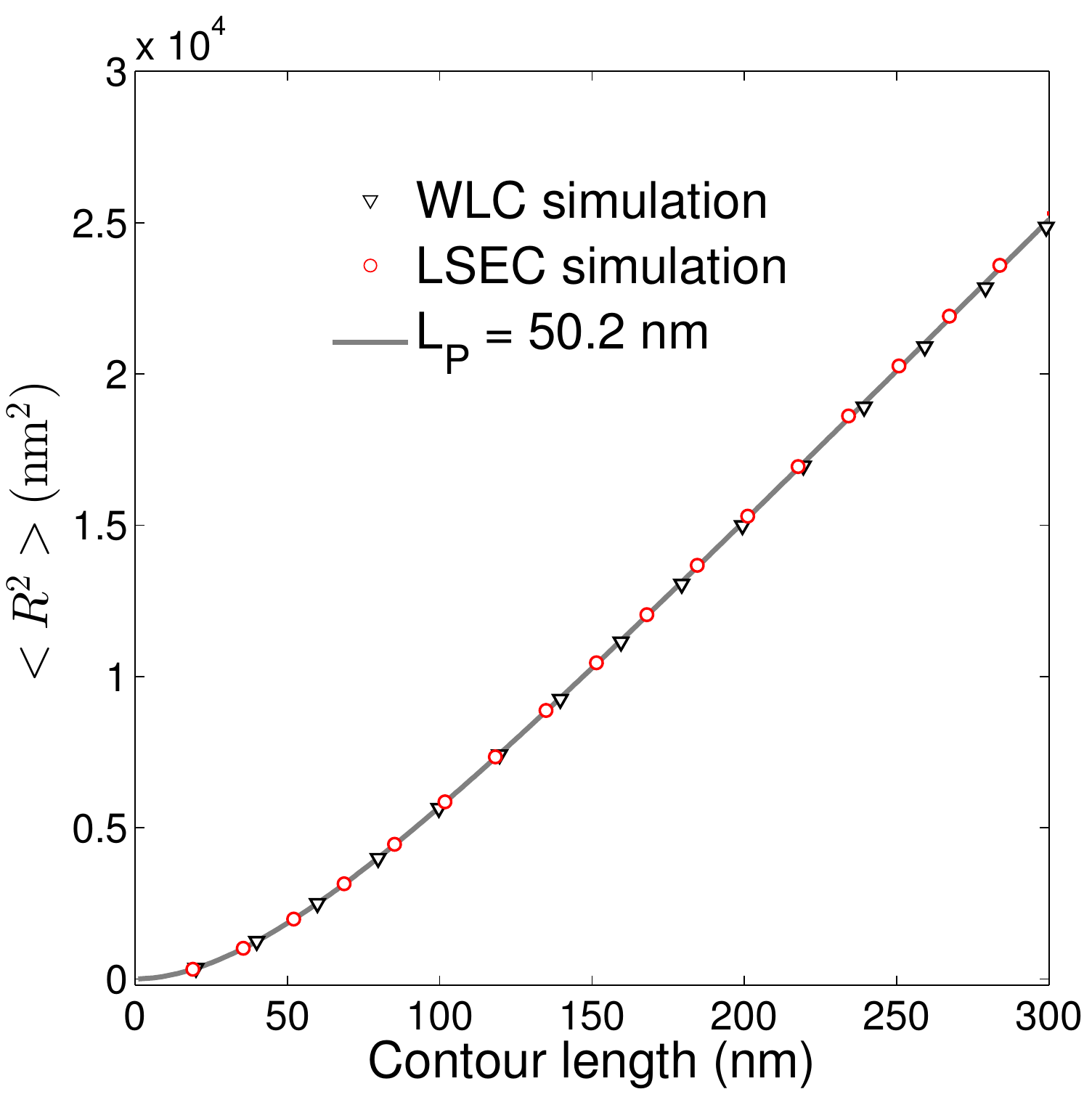}
\caption{\label{sfig2}{\bf The mean-square end-to-end distance as a function of DNA contour length.} We considered both WLC and LSEC models to generate chain conformations. The simulated mean-square end-to-end distances for WLC (black triangles) and LSEC (red circles) are compared against the analytical formula (Eq.˜\ref{r2}) for WLC with ${\sim}$50 nm persistence length (solid curve).}
\end{minipage}
\end{figure}

\begin{figure}
\begin{minipage}[c][\textheight]{\textwidth}
\includegraphics[width=12cm]{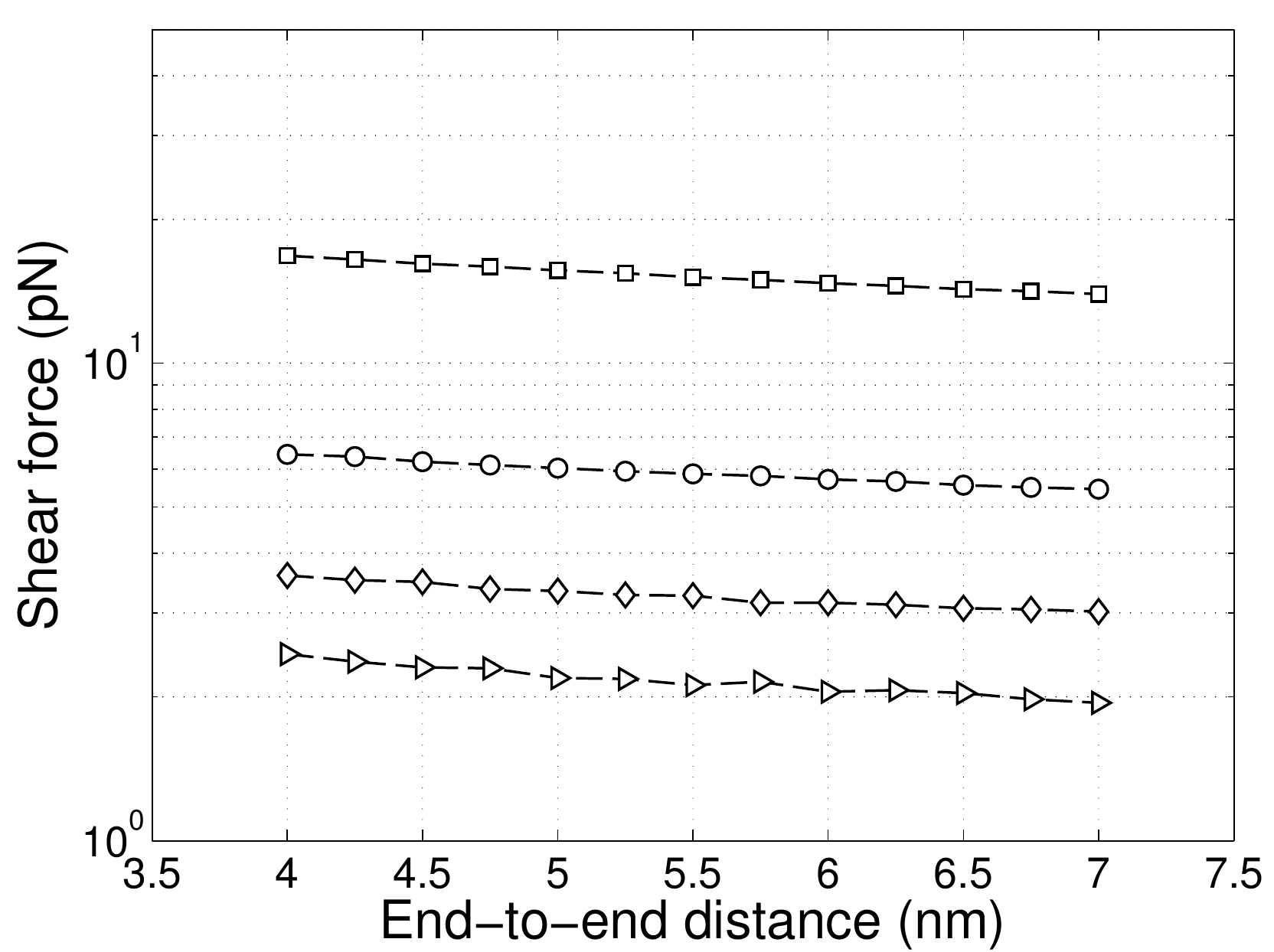}
\caption{\label{sfig3}{\bf Shear force vs. end-to-end distance.} The mean shear force was calculated from the WLC model for different loop sizes and different end-to-end distances ($r_0$). The shear forces at different loop sizes (square: 40 bp, circle: 70 bp, diamond: 100 bp, triangle: 130 bp) decrease only slightly as a function of the end-to-end distance. Since the linker duplex is extended by ${\sim}$1 nm before dissociation, our estimated force can be variable by ${\sim}$5\% for all loop sizes tested.}
\end{minipage}
\end{figure}

\begin{figure}
\begin{minipage}[c][\textheight]{\textwidth}
\includegraphics[width=12cm]{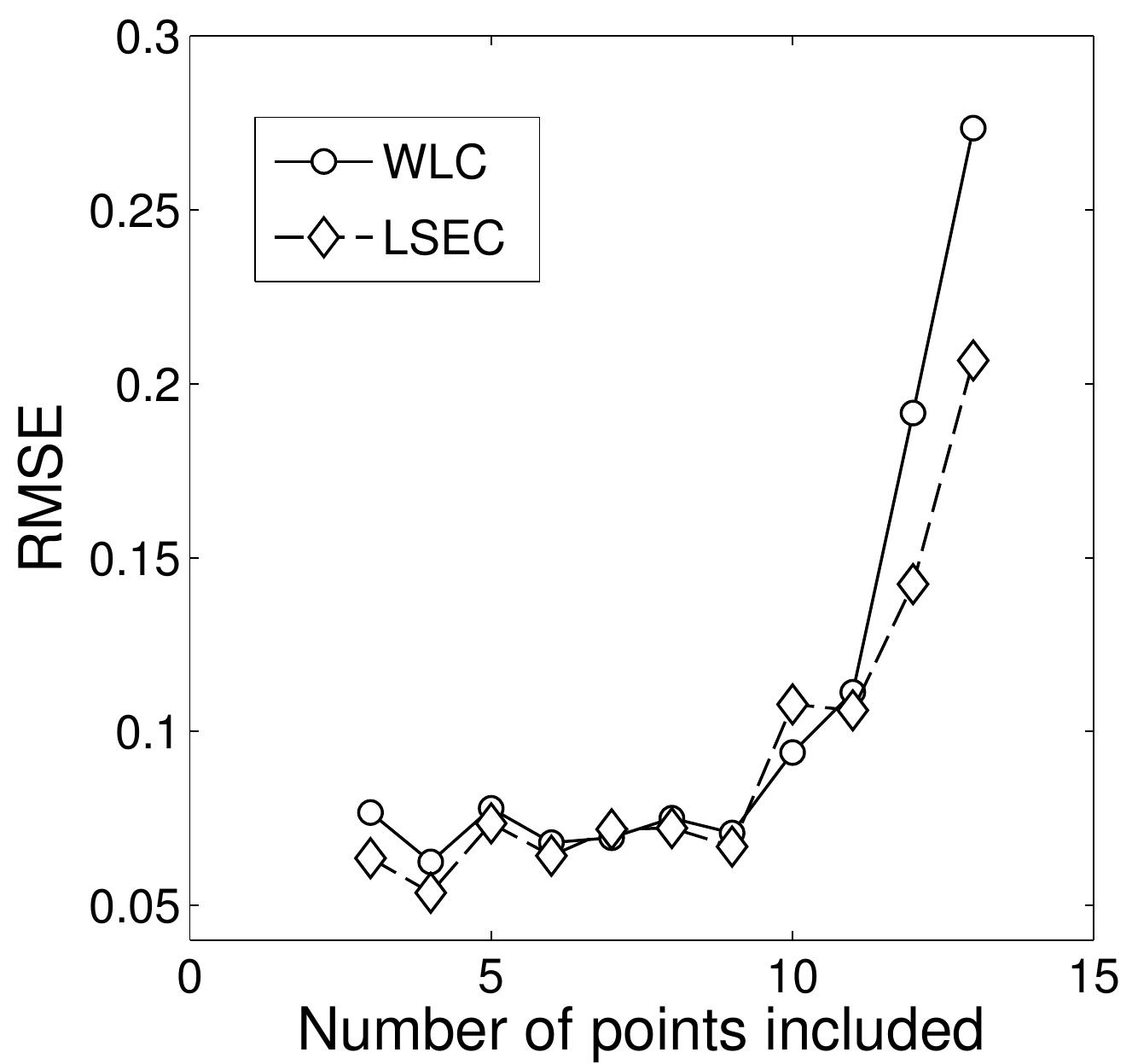}
\caption{\label{sfig4}{\bf RMSE analysis.} Root mean squared error (RMSE) of the linear regression of the logarithm of linker lifetime (in 50 mM NaCl) vs. shear force. The linear regression was performed with the `robustfit' function (MATLAB). To identify outliers, we compared the RMSE values resulting from different ranges of fitting.  For example, the last point is obtained when the entire range of 13 loop sizes from 189 bp down to the smallest 37 bp were included in the fitting. Including the last few points significantly increases the regression error, which indicates that the linear relationship predicted by Eq.˜\ref{bell} no longer holds for loop sizes smaller than 60 bp. Thus, we did not include three points corresponding to 37, 42, and 53 bp in the regression when extracting the fitting parameters, $\tau(0)$  and $\Delta r_0$.}
\end{minipage}
\end{figure}

To calculate $\langle R^2 \rangle$, one can generate a large set of chains using the Gaussian sampling method. For the WLC model, we chose the bending rigidity constant $k$ to be 73.53$k_B T$ for each 1-bp long monomer. For the LSEC model\cite{wiggins2006high,wiggins2006generalized}, we chose $B$ = 7.84$k_B T$ for each 7-bp long monomer (2.37 nm). The chosen parameters all predict a persistence length of ${\sim}$50 nm at large length scales (FIG. ~\ref{sfig2}).
\subsection*{J factor}
The looping probability ($P_1$) is the colocalization probability of two reactive ends of the same polymer within a small reaction volume $\delta V$. We do not know \textit{a priori} what $\delta V$ is, but it should be small enough to allow for the two ends of the polymer to react. Therefore, for cyclization of dsDNA with complementary single-stranded overhangs, its dimension should be on the order of the length of the single-stranded overhang (${\sim}$5 nm). The J factor is the effective concentration of one freely diffusing reactive end around the other that would give rise to the same colocalization probability, and can be determined without knowledge of $\delta V$. 

Without losing generality, we can fix one reactive end inside $\delta V$ and let other reactive ends freely diffuse at a molar concentration of $[X]$. The rate of a reactant diffusing into $\delta V$ is proportional to $[X]$ ($k_{in} [X]$) whereas the rate of the reactant diffusing out of the volume is concentration-independent ($k_{out}$). In typical aqueous reactions, the diffusive encounter between the two ends is much slower than the diffusive separation ($k_{out} \gg k_{in} [X]$)\cite{wang1966probability}. The equilibrium probability of intermolecular colocalization ($P_2$) is a function of $[X]$:
\begin{equation}\label{dimer} P_2([X])=\frac{k_{in} [X]}{k_{in} [X]+k_{out}}\approx\frac{k_{in}}{k_{out}} [X] \end{equation}
Therefore, the J factor is defined by 
\begin{equation}\label{jfactor} P_1=P_2(J)\,\,\,\,\therefore J=P_1 \frac{k_{out}}{k_{in}} \end{equation}
The J factor can be determined by measuring both intramolecular and intermolecular reaction kinetics. Both reactions follow a three-state reaction kinetics scheme: 
\begin{equation} a \rightleftharpoons b \rightarrow c.\end{equation}
Here, $b$ is the state of end-to-end colocalization without interaction, and $c$ is the high-FRET state stabilized by end-to-end annealing. If $k_{b\rightarrow c} \ll k_{b\rightarrow a}$, the apparent rate of $c$ formation ($k_c$) is proportional to the equilibrium probability of state $b$: 
\begin{equation}  k_c\approx\frac{k_{a\rightarrow b}}{k_{a\rightarrow b}+k_{b\rightarrow a}}k_{b\rightarrow c}= P_b k_{b\rightarrow c}\end{equation}

We denote the rate of annealing ($k_{b\rightarrow c}$) as $f_1$ for looping and $f_2$ for dimerization. The apparent looping rate ($k_1$) is 
\begin{equation}k_1 \approx P_1 f_1,\end{equation}
and the apparent dimerization rate ($k_2$) is
\begin{equation}k_2 \approx P_2 f_2\approx\frac{k_{in}}{k_{out}} [X] f_2\end{equation}
where we used Eq.˜\ref{dimer}. The second-order rate constant $k_2/[X]$ is usually referred to as the annealing rate constant in most other studies\cite{gao2006secondary,cisse2012rule,rauzan2013kinetics}. According to Eq.˜\ref{jfactor}, the J factor is related to the apparent rates by 
\begin{equation}  J=\frac{k_1}{k_2/[X]}\cdot\frac{f_2}{f_1}\end{equation}
Therefore, only if $f_1=f_2$ can we determine the J factor from the apparent rates in an unbiased manner.

For looping of long dsDNA, $f_1=f_2$ is generally accepted\cite{wang1966probability}. For looping of short dsDNA, however, $f_1 = f_2$ may be violated. In dimerization, the two ends approach each other from all 4$\pi$ steradians. In many of these colocalization events, the sticky ends are not optimally aligned for annealing. In looping, the reactive ends approach each other at a much narrower range of angles. As a result, the dangling overhangs with intrastrand stacking\cite{ohmichi2002long} may find each other in an anti-parallel orientation more often than in free diffusion. Hence, the entropic barrier for $f_1$ would be lower than for $f_2$. This effect is conceptually similar to rate enhancement in intramolecular reactions that far exceeds local concentration effect due to entropy\cite{page1971entropic} or orientation-dependent reactivity\cite{dafforn1971sensitivity,mazor1990effective}. 

\subsection*{Calcuating the free energy of kink formation}
We adopted the computational method in  \cite{zheng2009theoretical}, which is also conceptually similar to a more theoretical approach\cite{sivak2012consequences}. We considered the dinucleotide bending energy $E(\theta)$ with both the elastic bending term and kinking term using the functional form in Eq.~\ref{kink}. The critical kink angle ($\beta$) was defined as the intercept of the two terms. The equilibrium probability density ($p(\theta)$) or the partition function of the bending angle $\theta$ is proportional to the multiplicity of $\sin(\theta)$ and the Boltzmann factor
\begin{equation}\label{pa} p(\theta) \sim \sin(\theta) \exp(-E(\theta)/k_B T). \end{equation} 
The kinking probability ($P_k$) is the probability for $\theta$ to exceed the critical kink angle $\beta$, which is
\begin{equation}\label{pa} P_k = \frac{\int_{\beta}^{\pi}{\sin(\theta) \exp(-E(\theta)/k_B T) d\theta}}{\int_{0}^{\pi}{\sin(\theta) \exp(-E(\theta)/k_B T) d\theta}}. \end{equation}
The free energy of kink formation $\Delta G_k$ can be directly calculated from $P_k$ as $\Delta G_k = -k_B T \log(P_k)$. For example, if we consider an energy function $h + (\theta -b)^6$ with $h=15 k_B T$ and $b = 0.3$ for kink formation, $\Delta G_k = 10.6 k_B T$. If we assume no additional energy cost of kinking\cite{wiggins2005exact}, we have a little lower $\Delta G_k$ of $9.4 k_B T$, as expected. 

\begin{figure}
\begin{minipage}[c][\textheight]{\textwidth}
\includegraphics[height=12cm]{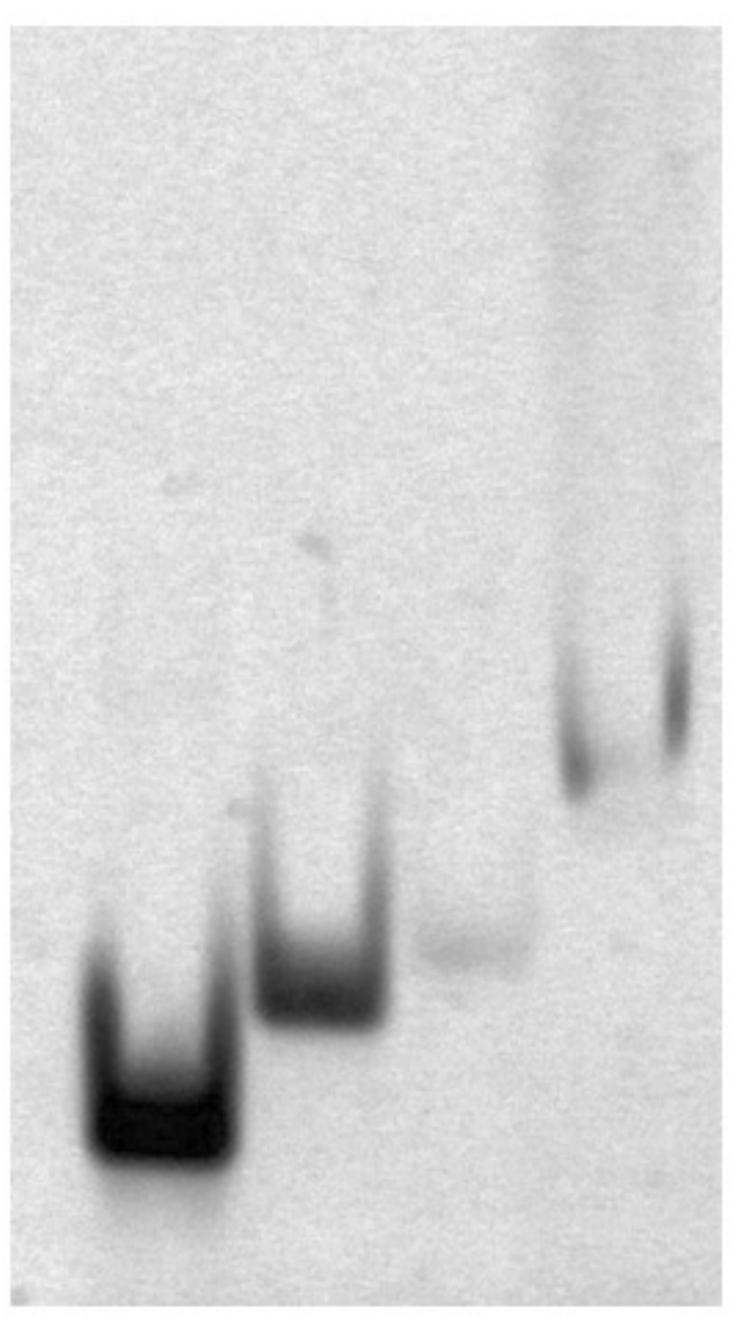}
\caption{\label{sfig5}{\bf Polyacrylamide gel image of the hybridized oligos.} From left to right, primer 1 ($5'$-/Cy3/GGTAAATTCACTAT CAACAACGAGCTCGAATG) only, 1:1 mixture of primer 1 and a blocking oligo ($5'$-CATTCGAGCTCGTTGTTG), primer 2 ($5'$-/BioTEG/GAAACATAG/iCy5/GAATTTACCGTGCCAGCAACAGATAGC) only, Lane 4: 1:1 mixture of primer 2 and a blocking oligo ($5'$-GCTATCTGTTGCTGGCAC).}
\end{minipage}
\end{figure}

\begin{figure}
\begin{minipage}[c][\textheight]{\textwidth}
\includegraphics[width=12cm]{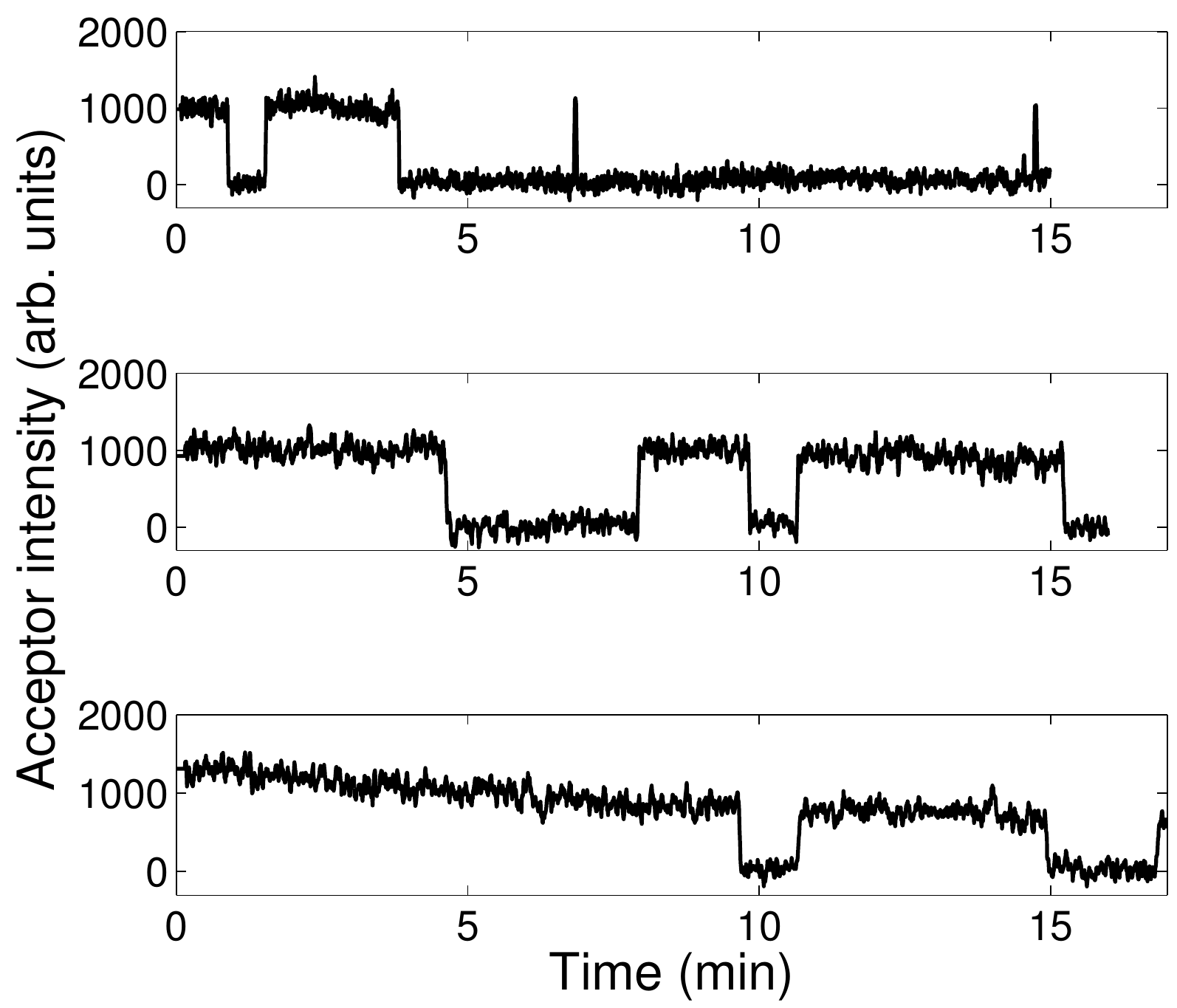}
\caption{\label{sfig6}{\bf Measuring $\tau(0)$.} Typical time
traces of reversible linker formation and separation in 50, 100 and 200 mM
[Na\textsuperscript{+}] (from top to bottom). Linker formation results in a burst in Cy5
intensity due to FRET. The survival probability of the dimer since $t=0$ is fitted with
a single exponential function to extract the linker lifetime at zero force
$\tau(0)$. The concentration of the free monomer was adjusted to obtain similar binding rates at different [Na\textsuperscript{+}]. }
\end{minipage}
\end{figure}

\begin{figure}
\begin{minipage}[c][\textheight]{\textwidth}
\includegraphics[width=12cm]{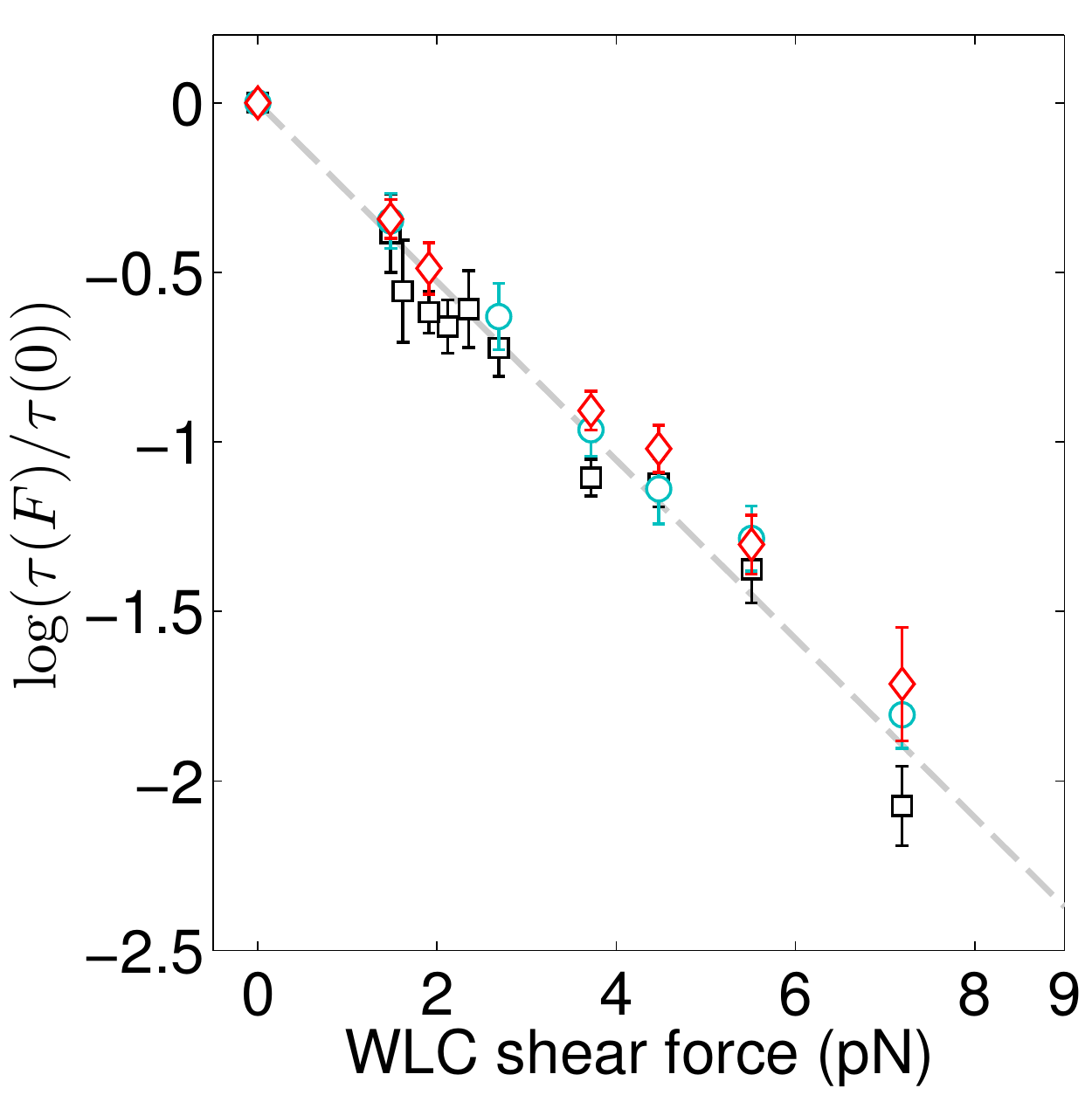}
\caption{\label{sfig7}{\bf Normalized lifetime vs. force} The looped state lifetimes $\tau$ measured at different sodium concentrations are normalized by their respective zero-force lifetimes $\tau(0)$. Black squares: 50 mM NaCl, blue circles: 100 mM NaCl and red diamonds: 200 mM NaCl.}
\end{minipage}
\end{figure}

\begin{figure}
\begin{minipage}[c][\textheight]{\textwidth}
\includegraphics[width=12cm]{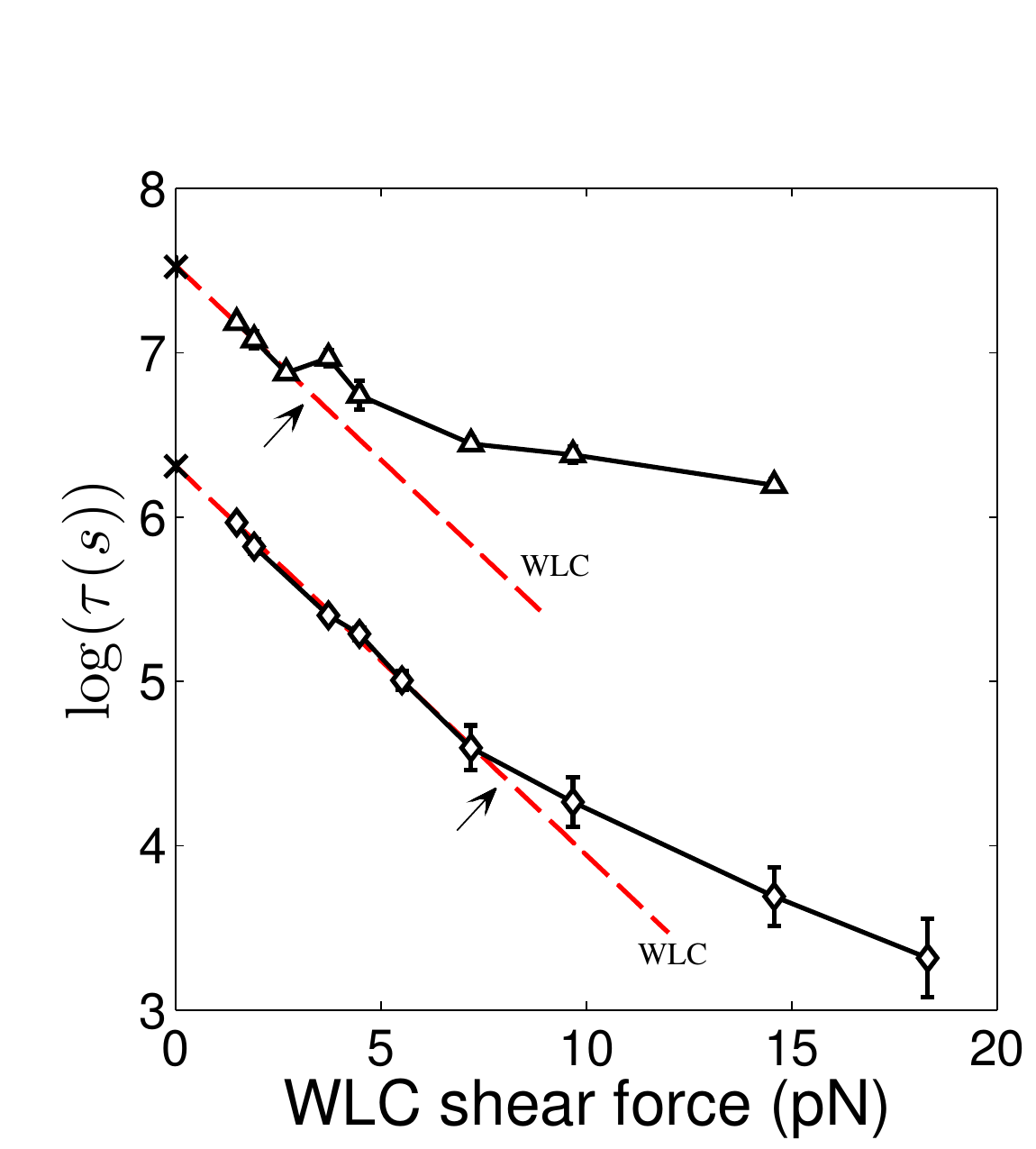}
\caption{\label{sfig8}{\bf Effect of salt on strong dsDNA bending.} Line scatter plot of the Loop lifetimes in 200 mM [Na\textsuperscript{+}] (diamonds) and 5 mM [Mg\textsuperscript{2+}] (triangles) are plotted against the shear forces predicted from the WLC model. Also shown are the zero-force lifetimes of the linker in respective salt conditions (marked `$\times$'). Dashed red curves represent the expected lifetimes based on a universal WLC model. Arrows indicate transition points at which the bending behavior of dsDNA deviates from the WLC.}
\end{minipage}
\end{figure}

\begin{figure}
\begin{minipage}[c][\textheight]{\textwidth}
\includegraphics[width=12cm]{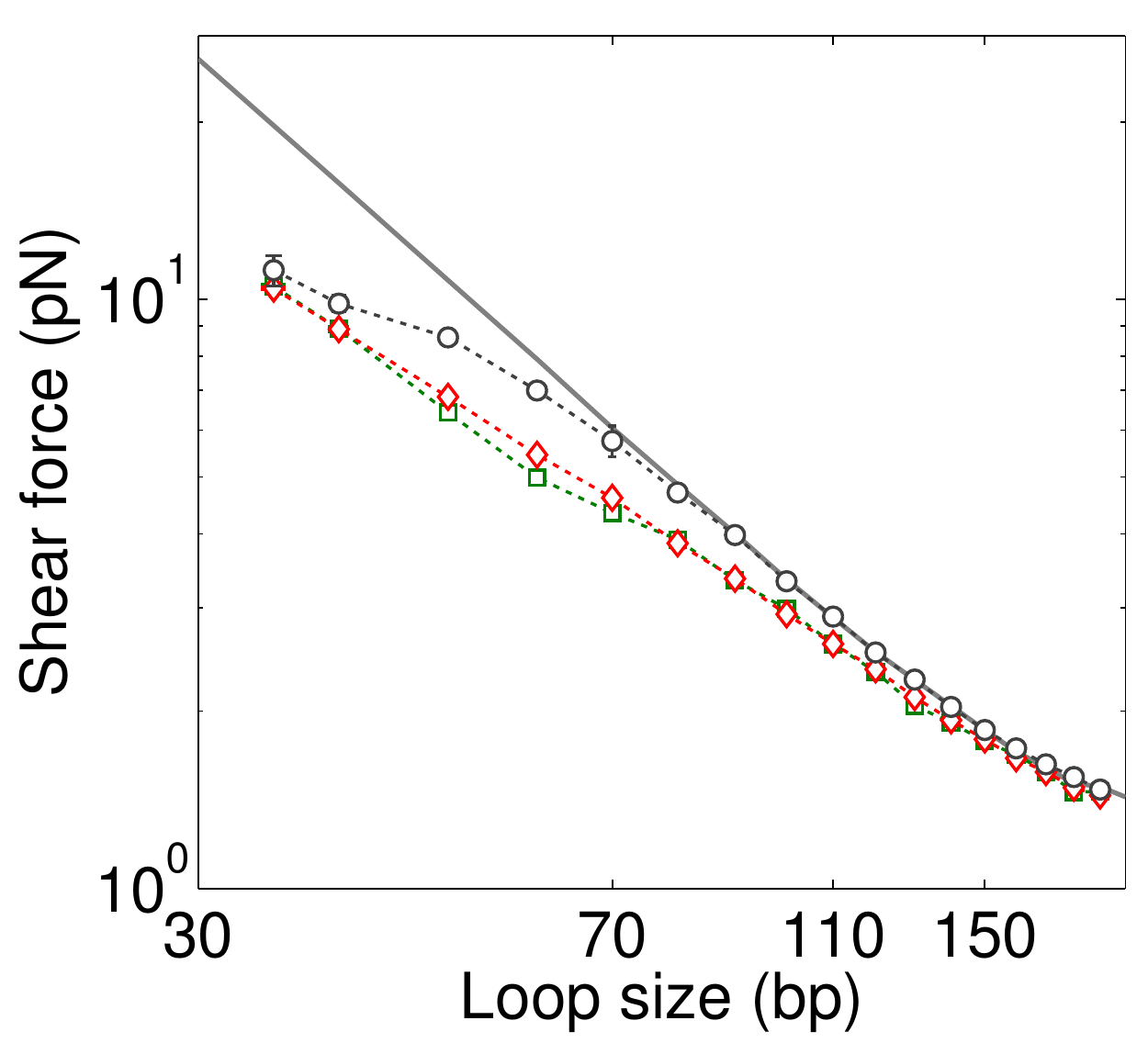}
\caption{\label{sfig9}{\bf Shear force from the KWLC model.} Shear forces are calculated from the KWLC model at different loop sizes with different parameters h and b (black circles: h = 22$k_B T$, b = 0.3; black squares: h = 13.64$k_B T$, b = 0; red diamonds: h = 10.56$k_B T$, b = -0.3). The gray curve represents the relationship calculated from the WLC model. Parameters h and b were chosen so that the predicted shear forces for loops below 50 bp are similar.}
\end{minipage}
\end{figure}

\begin{figure}
\begin{minipage}[c][\textheight]{\textwidth}
\includegraphics[width=12cm]{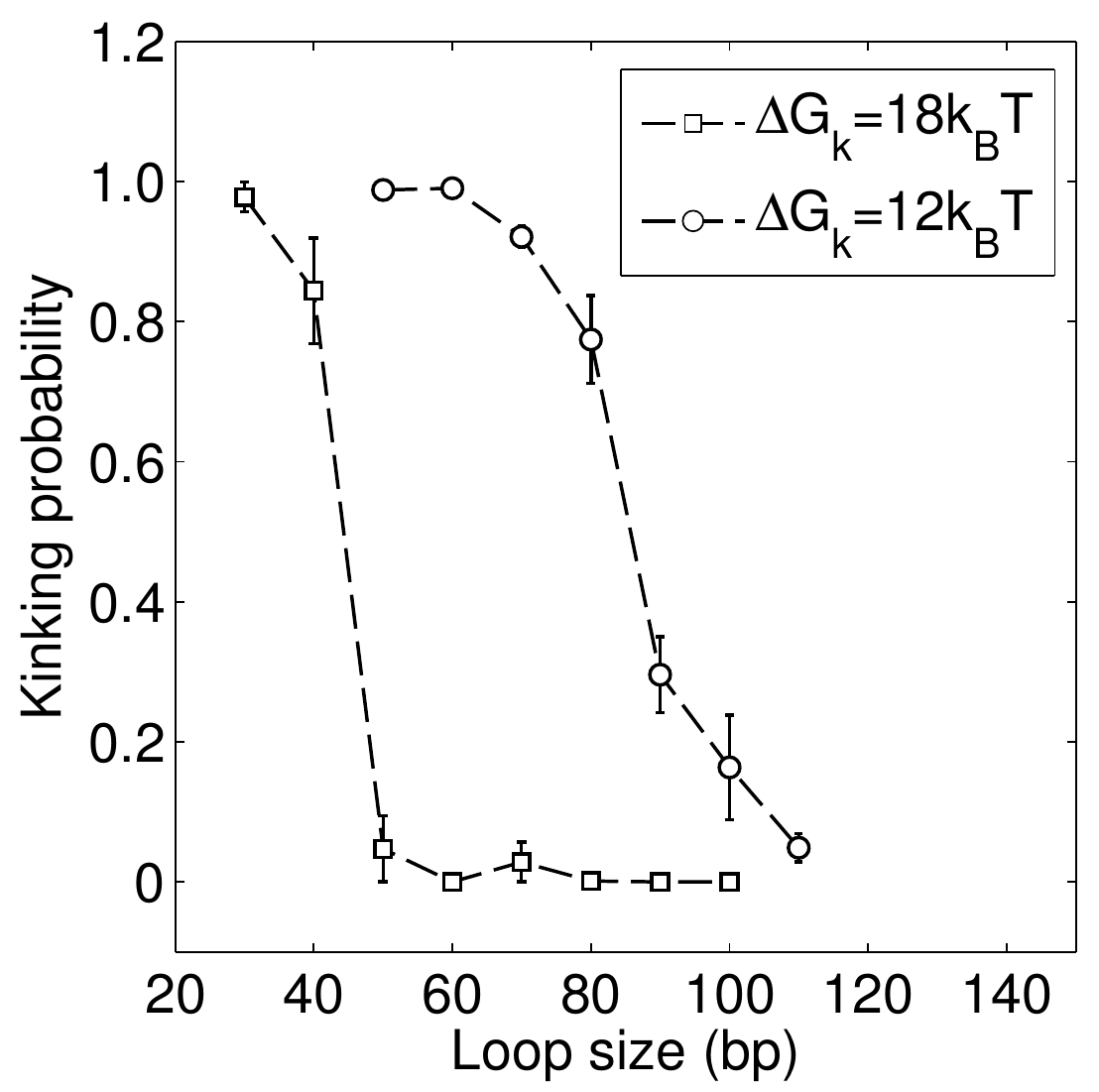}
\caption{\label{sfig10}{\bf Kinking probability in DNA miniciecles.} The kinking probabilities of DNA minicircles were calculated as a function of loop size using the KWLC model with different free energies of kink formation (squares: $\Delta G_k = 18 k_B T$ ($h=22 k_B T$, $b = 0.3$), circles: $\Delta G_k = 12 k_B T$ ($h = 17 k_B T$, $b = 0.7$). The SEM error bar for each loop size was calculated from 5 simulations.}
\end{minipage}
\end{figure}

\begin{figure}
\begin{minipage}[c][\textheight]{\textwidth}
\includegraphics[width=12cm]{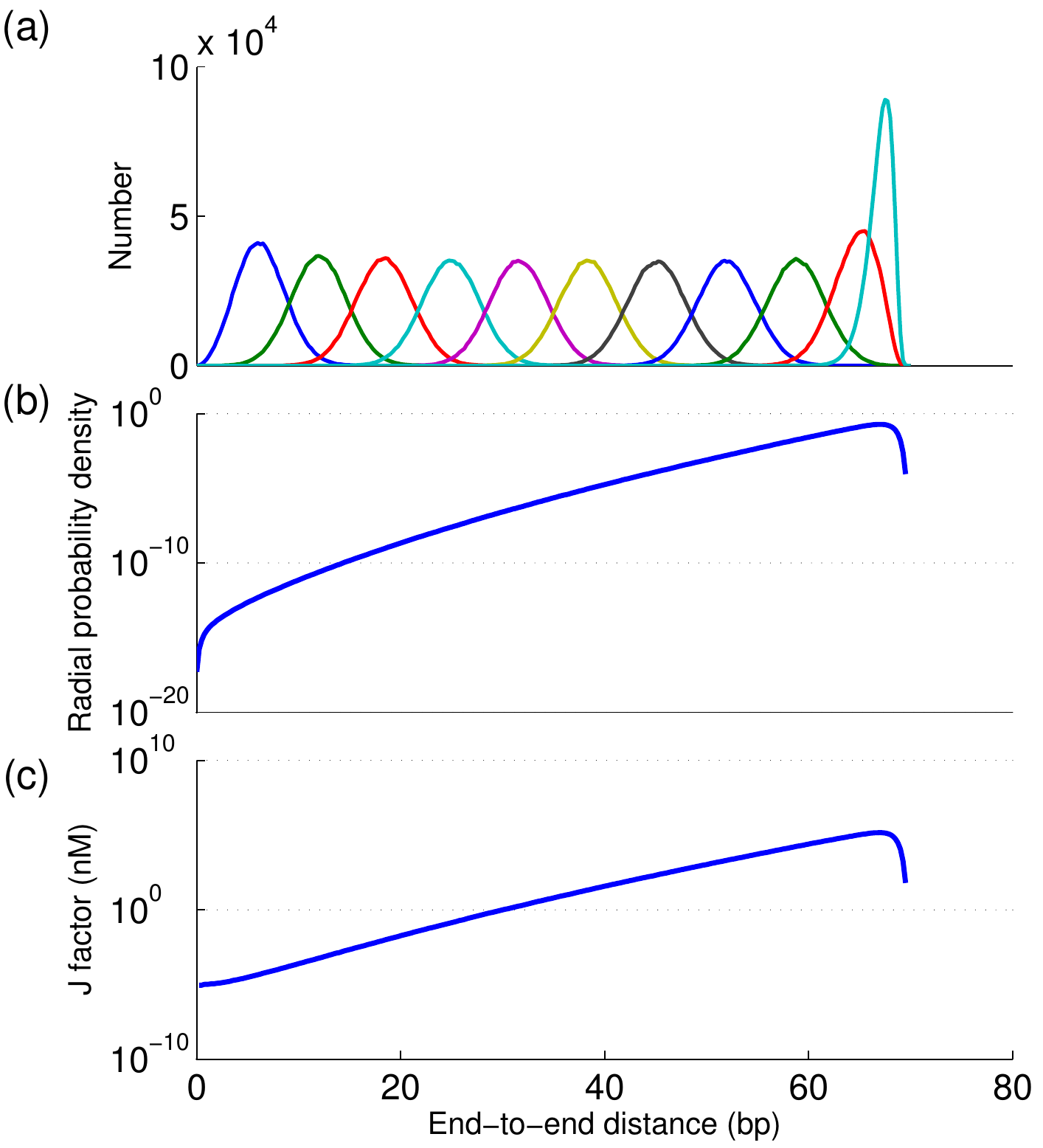}
\caption{\label{sfig12}{\bf J factor calculation by the weighted histogram analysis method.} (a) Umbrella sampling was performed at every 10-bp step. The spring constant was chosen so that neighboring histograms overlap significantly. Each histogram was obtained from $10^6$ MC conformations after 100,000 thermalization steps. (b) The radial probability distribution was obtained by iterating through Eq.˜\ref{wham}. (c) The J factor in nanomolar units can be obtained by dividing the amplitude of the radial probability distribution by $4\pi r^2 \Delta r$ and multiplying by $4.24 \times 10^{10}$.}
\end{minipage}
\end{figure}

\clearpage

\end{document}